\documentclass[twocolumn,aps,prl,showpacs,superscriptaddress]{revtex4}
\usepackage[dvips]{graphicx}
\usepackage{bm}
\begin{document}
\title{Local Spin Anisotropy Effects upon the Magnetization and Specific Heat of Dimer Single Molecule Magnets}
\author{Dmitri V. Efremov}
\email{efremov@theory.phy.tu-dresden.de} \affiliation{Institut
f{\"u}r Theoretische Physik, Technische Universit{\"a}t Dresden,
01062 Dresden, Germany}\author{Richard A. Klemm}
\email{klemm@phys.ksu.edu} \affiliation{Department of Physics,
Kansas State University, Manhattan, KS 66506 USA}
\date{\today}
\begin{abstract}
We  present an exactly solvable model of  equal spin $s_1$ dimer
single molecule magnets. The spins within each dimer interact
 via the Heisenberg and the most general quadratic global and
 local (single-ion) anisotropic spin  interactions,  and with the magnetic induction ${\bf B}$.
For  antiferromagnetic couplings and $s_1>1/2$,
 the  low temperature $T$ magnetization ${\bm M}({\bm B})$ exhibits $2s_1$ steps
  of universal height and midpoint slope, the $s$th step of which occurs at the
  non-universal level-crossing magnetic induction $B_{s,s_1}^{\rm
  lc}(\theta,\phi)$, where
  $\theta,\phi$ define the direction of  ${\bm B}$.  The specific
  heat  $C_V$ exhibits zeroes as $T\rightarrow0$ at these $B_{s,s_1}^{\rm
  lc}(\theta,\phi)$ values, which are equally surrounded by
  universal peak pairs as $T\rightarrow0$.  The non-universal $B_{s,s_1}^{\rm
  lc}(\theta,\phi)$ values lead to a rich variety of magnetization
  plateau
  behavior,
 the structure and anisotropy of which  depend
upon the various global and local anisotropic spin interaction
energies. We solve the model exactly for $s_1=1/2$, 1, and 5/2,
and present ${\bm M}({\bm B})$  and $C_V({\bm B})$ curves at low
$T$ for these cases.  For weakly anisotropic dimers, rather simple
analytic formulas for ${\bm M}({\bm B})$ and $C_V({\bm B})$ at
arbitrary $s_1$ accurately fit the exact solutions at sufficiently
low $T$ or large $B$. An expression for
 $B_{s,s_1}^{\rm lc}(\theta,\phi)$
accurate to second order in  the four independent anisotropy
energies is derived. Our results  are discussed with regard to
existing experiments on $s_1=5/2$ Fe$_2$ dimers, suggesting
further experiments on single crystals of these and some $s_1=9/2$
[Mn$_4$]$_2$ dimers are warranted.
\end{abstract}

 \vskip0pt\vskip0pt
\maketitle

\section{Introduction}

Single molecule magnets (SMM's) have been under intense study
recently, due to their potential uses in magnetic storage and
quantum computing.\cite{background,sarachik,loss}  The materials
consist of insulating crystalline arrays of identical SMM's  1-3
nm in size, each containing two or more magnetic ions. Since the
magnetic ions in each SMM are surrounded by non-magnetic ligands,
the intermolecular magnetic interactions are usually negligible.
Although the most commonly studied SMM's are the high-spin
Mn$_{12}$ and Fe$_8$,\cite{background,sarachik,loss,Fe8,WS}  such
SMM's contain a variety of ferromagnetic (FM) and
antiferromagnetic (AFM) intramolecular interactions, rendering
unique fits to a variety of experiments difficult.\cite{Fe8spin9}

In addition, there have been many studies of AFM Fe$_n$ ring
compounds, where $n=6, 8, 10, 12,$
etc.\cite{Fering1,Fe6ring,Fe8ring,Fering2}  In these studies,
analyses of inelastic neutron diffraction data and
 the magnetic induction ${\bm B}$ dependence of the
low-temperature $T$ specific heat and magnetization steps were
made, using the isotropic Heisenberg near-neighbor exchange
interaction, the Zeeman interaction, and various near-neighbor
spin anisotropy
interactions.\cite{Fering1,Fe6ring,Fe8ring,Fering2} However, the
rings were so complicated that  analyses of the data using those
simple models were inaccessible to present day
computers.\cite{Fe6ring,Fe8ring} Thus, those authors used either
simulations or phenomenological fits to a first-order perturbation
expansion with different spin anisotropy values for each global
ring spin value.\cite{Fering2,Fe6ring,Fe8ring}

Here we focus on the simpler cases of equal spin $s_1=s_2$
magnetic dimers, for which the full spin anisotropy effects can be
evaluated analytically,  investigated in detail numerically, and
compared with experiment.
 AFM dimers with  $s_1=1/2, 3/2$,\cite{V2neutron,Gudel,V2P2O9,ek} and various
forms of Fe$_2$ with $s_1=5/2$
 were
 studied recently.\cite{Fe2,Fe2mag,Fe2Cl,taft,Fe2Cl3,Fe2Clnew} Several
  Fe$_2$ dimers and effective
$s_1=9/2$ dimers  of the type
[Mn$_4$]$_2$,\cite{Mn4dimer,Mn4dimerDalal} have magnetic
interactions weak enough that their effects can be probed at
$T\approx$ 1K with presently available  ${\bm B}$. A comparison of
our  results with magnetization ${\bm M}$ versus ${\bm B}$ step
data on a Fe$_2$ dimer
 strongly suggests  a
substantial presence  of local spin anisotropy.\cite{Fe2Cl}

The paper is organized as follows.  In Section II, we present the
model in the crystal representation and given exact formulas for
the matrix elements.  The general thermodynamics are presented in
Section II, along with the universal behavior of the $M(B)$ and
$C_V(B)$ behavior associated with the energy level crossing. In
Section IV, we solve the model exactly for $s_1=1/2$, giving
analytic expressions for the magnetization and specific heat. In
Section V, we discuss the exact solution for $s_1=1$, present the
equations from which the eigenvalues are readily obtained, and
give numerical examples of the low-$T$ magnetization and specific
heat curves.  In Section VI, we present numerical examples of the
low-$T$ $s_1=5/2$ magnetization and specific heat curves. In
section VII, we rotate to the induction representation, and give
the eigenstates to first order in the anisotropy energies.  These
are used to obtain asymptotic expressions for the magnetization
and specific heat for arbitrary $s_1$ that are highly accurate at
sufficiently low $T$ and/or large $B$.  In addition, analytic
formulas for the magnetic inductions at which the level crossings
occur are provided, accurate to second order in the anisotropy
energies. Finally, in Section VIII we discuss our results with
regard to experiments on Fe$_2$ dimers.

\section{The Model in the Crystal Representation}

 We represent the $s_1=s_2$ dimer
quantum states, $|\psi_s^m\rangle$ in terms of the global (total)
spin and magnetic quantum numbers $s$ and $m$, where ${\bm S}={\bm
S}_1+{\bm S}_2$ and $S_z={\bm S}\cdot\hat{\bm z}$ satisfy ${\bm
S}^2|\psi_s^m\rangle=s(s+1)|\psi_s^m\rangle$ and
$S_z|\psi_s^m\rangle=m|\psi_s^m\rangle$, where
$s=0,1,\ldots,2s_1$, $m=-s,\ldots, s$, and we  set $\hbar=1$.
 We also have $S_{\pm}|\psi_s^m\rangle=A_s^{\pm m}|\psi_s^{m\pm1}\rangle$, where
$S_{\pm}=S_x\pm iS_y$ and
\begin{eqnarray}
A_s^m&=&\sqrt{(s-m)(s+m+1)}.\label{Asm}
\end{eqnarray} For an arbitrary  ${\bm B}$,  we assume the Hamiltonian has the
form ${\cal H}={\cal H}_0+{\cal H}_a+{\cal H}_b+{\cal H}_d+{\cal
H}_e$, where
\begin{eqnarray}{\cal H}_0&=&-J{\bm
S}^2/2-\gamma{\bm S}\cdot{\bm B} \end{eqnarray}
 contains the
Heisenberg exchange and Zeeman interactions, the gyromagnetic
ratio $\gamma=g\mu_B$, where $g\approx2$ and $\mu_B$ is the Bohr
magneton. The global axial and azimuthal anisotropy terms
\begin{eqnarray}{\cal H}_b&=&-J_bS_z^2\label{Hb} \end{eqnarray}
and
\begin{eqnarray}{\cal H}_d&=&-J_d(S_x^2-S_y^2),\label{Hd}
\end{eqnarray} respectively, only involve components of ${\bm S}$, but have been the main
anisotropy terms discussed in the SMM
literature,\cite{WS,Mn4dimer} so we have included them for
comparison. Such terms have been commonly studied as an effective
Hamiltonian for a singlet orbital ground state, in which the
tensor global spin interaction  with a fixed spin quantum number
$s$ has the form ${\bm S}\cdot\tensor{\bm\Lambda}\cdot{\bm S}$,
resulting in the principal axes $x,y$ and $z$.\cite{WaldmannNi}
For a dimer, we take $\tensor{\bm \Lambda}$ to be diagonal in the
orientation pictured in Fig. 1. Then
$J=-(\Lambda_{xx}+\Lambda_{yy})/2$,
$J_b=-\Lambda_{zz}+(\Lambda_{xx}+\Lambda_{yy})/2$, and
$J_d=(\Lambda_{yy}-\Lambda_{xx})/2$.  Taking $|J_d/J|\ll1$ and
$|J_b/J|\ll1$ still leaves $J_d/J_b$ unrestricted.  The single-ion
axial and azimuthal anisotropy terms,
\begin{eqnarray}
{\cal H}_a&=&-J_a\sum_{i=1}^2S_{iz}^2\label{lz}\\
 \noalign{\rm and} {\cal
H}_e&=&-J_e\sum_{i=1}^2\Bigl(S_{ix}^2-S_{iy}^2\Bigr),\label{le}
\end{eqnarray}
respectively, arise from spin-orbit interactions of the local
crystal field with the individual spins. These terms have usually
been  neglected in the SMM literature, but have been studied with
regard to complexes containing a single magnetic ion, such as
Ni$^{+2}$,\cite{WaldmannNi} and with regard to clusters of larger
numbers of identical magnetic ions.\cite{WaldmannNi,Almenar}

The local axially and azimuthally anisotropic exchange
interactions
\begin{eqnarray} {\cal
H}_f&=&-J_fS_{1z}S_{2z}\label{Hf},\\
{\cal H}_c&=&-J_c(S_{1x}S_{2x}-S_{1y}S_{2y}),\label{Hc}
\end{eqnarray}
 satisfy
 \begin{eqnarray}
 2{\cal H}_f/J_f&=&{\cal
H}_b/J_b-{\cal H}_a/J_a,\label{Hfequiv}\\
2{\cal H}_c/J_c&=&{\cal H}_d/J_d-{\cal
 H}_e/J_e,\label{Hcequiv}
 \end{eqnarray}
so we need only include either either ${\cal H}_a$ or ${\cal H}_f$
and ${\cal H}_e$ or ${\cal H}_c$, respectively.\cite{WaldmannNi}
That is, if we stick to the Hamiltonian ${\cal H}$, we may
incorporate the effects of ${\cal H}_f$ and ${\cal H}_c$ by
letting $J_b\rightarrow J_b+J_f/2$, $J_a\rightarrow J_a-J_f/2$,
and $J_d\rightarrow J_d+J_c/2$, $J_e\rightarrow J_e-J_c/2$,
respectively. Since  ${\cal H}_a$ and ${\cal H}_e$ describe the
axial and azimuthal anisotropy each single ion attains from its
surrounding environment, they are the physically relevant local
anisotropy interactions.

For the case of Fe$_2$,\cite{Fe2} a constituent of the high-spin
SMM Fe$_8$ and the AFM Fe$_n$ rings,\cite{Fe8,Fe6ring,Fe8ring} the
exchange between the Fe$^{+3}$ $s_1=5/2$ spins occurs via two
oxygen ions, and these four ions essentially lie in the same
($xz$) plane.\cite{Fe2,taft} We set the $z$ axis parallel to the
dimer axis, as pictured in Fig. 1.  Since the quantization axis is
along the dimer axis, which is fixed in a crystal, we  denote this
representation as the crystal representation.

We generally expect each of the $J_j$ for $j=a,b,d,e$ to satisfy
$|J_i/J|\ll1$, but there are not generally any other restrictions
upon the various magnitudes of the $J_j$.  Since all dimers known
to date have predominantly AFM couplings ($J<0$), and also because
their magnetizations and specific heats are particularly
interesting, we shall only consider AFM dimers.  In addition,
since no studies on unequal-spin dimers have been reported to our
knowledge, we shall only treat the equal-spin $s_1=s_2$ case.  We
note that for equal spin dimers, the group symmetry of the dimer
environment is $C_{2v}$, so that Dzaloshinski{\u\i}-Moriya
interactions do not arise.\cite{WaldmannNi}  Hence, our
Hamiltonian ${\cal H}$ is the most general quadratic anisotropic
spin Hamiltonian of an equal-spin dimer.

\begin{figure}
\includegraphics[width=0.15\textwidth]{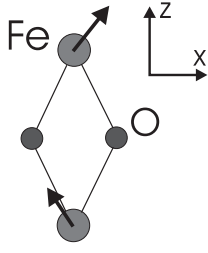}\vskip5pt
\caption{ Sketch of an Fe$_2$ dimer, with two bridging O$^{-2}$
ions (O). Ligands (not pictured) are
 attached to the  Fe$^{+3}$ ions (Fe).  The arrows signify
 spins.}\label{fig1}\end{figure}

For ${\bm B}=B(\sin\theta\cos\phi,\sin\theta\sin\phi,\cos\theta)$,
we have
\begin{eqnarray}
{\cal H}_0|\psi_s^m\rangle&=&E^m_{s}|\psi_s^m\rangle+\delta
E\sum_{\sigma=\pm1}e^{-i\sigma\phi}A_s^{\sigma
m}|\psi_s^{m+\sigma}\rangle,\nonumber\\
& &\\ {\cal
H}_b|\psi_s^m\rangle&=&-J_bm^2|\psi_s^m\rangle,\label{Hbpsi}\\
\noalign{\rm and}\nonumber\\ {\cal
H}_d|\psi_s^m\rangle&=&\frac{-J_d}{2}\sum_{\sigma=\pm1}F_s^{\sigma
m}|\psi_s^{m+2\sigma}\rangle,\label{Hdpsi} \\ \noalign{\rm
where}\nonumber\\
E^m_s&=&-Js(s+1)/2-mb\cos\theta,\\
\delta E&=&-\frac{1}{2}b\sin\theta, \\
b&=&\gamma B.\label{b}\\
\nonumber{\rm
and}\nonumber\\
F_{s}^{x}&=&A_s^{x}A_s^{1+x}.\label{Fsx} \end{eqnarray}

 ${\cal
H}_a$ and ${\cal H}_c$
 contain the individual spin operators $S_{iz}$ and
$S_{i\pm}$ for $i=1,2$. After some Clebsch-Gordon algebra
involving the Wigner-Eckart theorem, for arbitrary  $(s_1,s,m)$,
\begin{eqnarray}
S_{i\pm}|\psi_s^m\rangle&=&\frac{1}{2}A_s^{\pm
m}|\psi_s^{m\pm1}\rangle\mp\frac{1}{2}(-1)^{i}\Bigl(C_{s,s_1}^{\pm
m}|\psi_{s-1}^{m\pm1}\rangle\nonumber\\
& &\qquad-C_{s+1,s_1}^{-1\mp m}|\psi_{s+1}^{m\pm1}\rangle\Bigr),\label{sipm}\\
S_{iz}|\psi_s^m\rangle&=&\frac{m}{2}|\psi_s^m\rangle-\frac{1}{2}(-1)^{i}\Bigl(D_{s,s_1}^m|\psi_{s-1}^m\rangle\nonumber\\
& &\qquad +D_{s+1,s_1}^m|\psi_{s+1}^m\rangle\Bigr),\label{siz}
\end{eqnarray}
 where
 \begin{eqnarray}
 C_{s,s_1}^m&=&\sqrt{(s-m)(s-m-1)}\eta_{s,s_1},\label{Css1m}\\
  D_{s,
s_1}^m&=&\sqrt{(s^2-m^2)}\eta_{s,s_1},\label{Dss1m} \end{eqnarray}
and
\begin{eqnarray}
\eta_{s,s_1}&=&\sqrt{[(2s_1+1)^2-s^2]/(4s^2-1)}.\label{eta}
\end{eqnarray}
  For $s=0$, we require $m=0$, for which
$C_{0,s_1}^0=D_{0,s_1}^0=0$.  We then find
\begin{eqnarray}
{\cal
H}_a|\psi_s^m\rangle&=&\frac{-J_a}{2}\Bigl(G_{s,s_1}^m|\psi_s^m\rangle\nonumber\\
& &\qquad
+\sum_{\sigma'=\pm1}H_{s,s_1}^{m,\sigma'}|\psi_{s+2\sigma'}^m\rangle\Bigr),\\
{\cal
H}_e|\psi_s^m\rangle&=&-\frac{J_e}{4}\sum_{\sigma=\pm1}\Bigl(L_{s,s_1}^{\sigma
m}|\psi_s^{m+2\sigma}\rangle\nonumber\\
& &\qquad+\sum_{\sigma'=\pm1}K_{s,s_1}^{\sigma
m,\sigma'}|\psi_{s+2\sigma'}^{m+2\sigma}\rangle\Bigr),\label{Hepsi}\\
G_{s,s_1}^m&=&s(s+1)-1\nonumber\\
& &+[m^2+1-2s(s+1)]\alpha_{s,s_1},\label{Gss1m}\\
H_{s,s_1}^{m,\sigma'}&=&D_{s+(\sigma'+1)/2,s_1}^mD_{s+(3\sigma'+1)/2,s_1}^m,\label{Hss1msigma}\\
K_{s,s_1}^{x,\sigma'}&=&C_{s+(\sigma'+1)/2,s_1}^{-\sigma'x-(1+\sigma')/2}C_{s+(3\sigma'+1)/2,s_1}^{-\sigma'x-(3\sigma'+1)/2},\label{Kss1xsigma}\\
L_{s,s_1}^x&=&2F_s^x\alpha_{s,s_1},\label{Lss1}\\
\noalign{and}
\alpha_{s,s_1}&=&\frac{3s(s+1)-4s_1(s_1+1)-3}{(2s-1)(2s+3)}.\label{alphass1}
\end{eqnarray}
We note that $\alpha_{s,s_1}=1-\eta_{s,s_1}^2-\eta_{s+1,s_1}^2$
and that Eq. (\ref{alphass1}) holds for $s\ge0$. Equations
(\ref{sipm}) and (\ref{siz}) allow for an exact solution to the
most general Hamiltonian of arbitrary order in the individual spin
operators. In all previous treatments of more complicated spin
systems with similar anisotropic interactions, it was only
possible to obtain numerical solutions, and therefore the full
anisotropy of the magnetization and specific heat was not
calculated.\cite{WaldmannNi,Almenar}
 The operations of ${\cal H}_0$, ${\cal H}_b$ and
${\cal H}_d$  satisfy the selection rules $\Delta s=0$, $\Delta
m=0,\pm1,\pm2$.  The local anisotropy interactions ${\cal H}_a$
and ${\cal H}_e$
 allow transitions satisfying $\Delta s=0,\pm2$,
$\Delta m=0$, and  $\Delta s=0,\pm2$, $\Delta m=\pm2$,
respectively, so in the presence of either of these interactions,
$s$ is no longer a good quantum number, unless $s_1=1/2$.

\section{General Thermodynamics}

In order to obtain the thermodynamic properties, we first
calculate the canonical partition function, $Z={\rm
Tr}\exp(-\beta{\cal H})$. Since ${\cal H}$ is not diagonal in the
$(s,m)$ representation, we must construct the wave function from
all possible spin states. We then write
\begin{eqnarray}
Z&=&{\rm Tr}\langle \Psi_{s_1}|e^{-\beta{\cal
H}}|\Psi_{s_1}\rangle,
\end{eqnarray}
where $|\Psi_{s_1}$ is constructed from the $\{|\psi_s^m\rangle\}$
basis as
\begin{eqnarray}
\langle\Psi_{s_1}|&=&\Bigl(<\psi_{2s_1}^{2s_1}|,
\langle\psi_{2s_1}^{2s_1-1}|, \ldots, \langle\psi_1^0|,
\langle\psi_1^{-1}|, \langle\psi_0^0|\Big),\nonumber\\
\end{eqnarray}
where $\beta=1/(k_BT)$ and  $k_B$ is Boltzmann's constant. To
evaluate the trace, it is  useful to diagonalize the $\langle
\Psi_{s_1}|{\cal H}|\Psi_{s_1}\rangle$ matrix. To do so, we let
$|\Psi_{s_1}\rangle={\bm U}|\Phi_{s_1}\rangle$, where
\begin{eqnarray}
\langle\Phi_{s_1}|&=&\Bigl(\langle\phi_{n_{s_{1}}}|,\langle\phi_{n_{s_{1}}-1}|,\ldots,\langle\phi_1|\Bigr),
\end{eqnarray}
is constructed from the new orthonormal basis
$\{|\phi_n\rangle\}$, and ${\bm U}$ is a unitary matrix of rank
$n_{s_{1}}=(2s_1+1)^2$. Choosing ${\bm U}$ to diagonalize ${\cal
H}$, we generally obtain ${\cal
H}|\phi_n\rangle=\epsilon_n|\phi_n\rangle$ and the partition
function for a SMM dimer,
\begin{eqnarray}
Z&=&\sum_{n=1}^{n_{s_{1}}}\exp(-\beta\epsilon_n).
\end{eqnarray}
The specific heat $C_V=k_B\beta^2\partial^2\ln Z/\partial\beta^2$
is then easily found at all $T, {\bm B}$, \begin{eqnarray}
C_V&=&\frac{k_B\beta^2}{Z^2}\biggl[Z\sum_{n=1}^{n_{s_1}}\epsilon_n^2e^{-\beta\epsilon_n}
-\Bigl(\sum_{n=1}^{n_{s_1}}\epsilon_ne^{-\beta\epsilon_n}\Bigr)^2\biggr].\label{sh}
\end{eqnarray}
It is easily seen from Eq. (\ref{sh}) that at the induction
$B_{s,s_1}^{\rm lc}$ corresponding to the $s$th level crossing in
which $\epsilon_s=\epsilon_{s-1}$, $C_{V}\rightarrow0$ as
$T\rightarrow0$.
 The magnetization
\begin{eqnarray}
{\bm M}&=&\frac{1}{Z}\sum_{n=1}^{n_{s_{1}}}{\bm\nabla}_{\bm
B}(\epsilon_n)\exp(-\beta\epsilon_n),\label{mag}
\end{eqnarray}
 requires ${\bm\nabla}_{\bm
B}(\epsilon_n)$ for each ${\bm B}$.   As $T\rightarrow0$, at most
two eigenstates are relevant.  For most $B$ values, only one
$\epsilon_n$ is important.  But near the $s$th level-crossing
induction  $B_{s,s_1}^{\rm lc}(\theta,\phi)$ at which
$\epsilon_s=\epsilon_{s-1}$, two eigenstates are relevant.  We
then have
\begin{eqnarray} C_{V}\Bigl[
B^{\rm lc}_{s,s_1}(\theta,\phi)\Bigr]&{\rightarrow\atop{T\rightarrow0}}&0,\label{CVzeroes}\\
C_V\Bigl[B_{s,s_1}^{\rm lc}(\theta,\phi)\pm
\frac{2c}{\gamma\beta}\Bigr]&{\rightarrow\atop{k_BT\ll|J|}}&C_V^{\rm peak}\label{peak}\\
C_V^{\rm peak}/k_B&= &\Bigl(\frac{c}{\cosh
c}\Bigr)^2\nonumber\\
& \approx&0.439229,\label{peakheight}\\ \noalign{\rm and}
M\Bigl[B_{s,s_1}^{\rm
lc}(\theta,\phi)\Bigr]/\gamma&{\rightarrow\atop{T\rightarrow0}}&
s-\frac{1}{2},\label{Msteps}\\
\frac{dM}{\gamma dB}\Bigr|_{B_{s,s_1}^{\rm
lc}(\theta,\phi)}&=&\frac{\beta\gamma}{4},\label{slope}
\end{eqnarray}  where $s=1,\ldots,2s_1$ and $c\approx 1.19967864$ is the solution to $\tanh c=1/c$.
The easiest way to obtain Eq. (\ref{Msteps}) is  to first rotate
the crystal so the quantization axis is along ${\bm B}$, as
discussed in Appendix B. We note that $B_{s,s_1}^{\rm
lc}(\theta,\phi)$ depends upon $s, s_1$, and the direction of
${\bm B}$ when anisotropic interactions are present. Hence, the
heights and midpoint slopes  of the $2s_1$ $M(B)$ steps are
universal, but the step positions and hence their plateaus are
not. Correspondingly, the $2s_1$ positions of the $C_V(B)$ zeroes
are non-universally spaced, but each zero is equally surrounded by
two peaks of equal, universal height, the universal spacing
between which is $\propto T$ as $T\rightarrow0$.  Hence, the
non-universal level-crossing inductions $B_{s,s_1}^{\rm
lc}(\theta,\phi)$ fully determine the low-$T$ thermodynamics of
AFM dimers.  In the next three sections, we consider the special
cases of $s_1=1/2,1$ and 5/2.  Then, in Section VII and Appendix
D, we present out general expression for $B_{s,s_1}^{\rm
lc}(\theta,\phi)$ accurate to second order in each of the $J_j$.
We remark that a double peak in the low-$T$ $C_V(B)$ curve has
been seen experimentally in a much more complicated Fe$_6$ ring
compound, and was attributed to level crossing.\cite{Fe6ring}

\section{Analytic results for spin 1/2}

Plots of $C_V/k_B$ and  $M/\gamma$ versus $\gamma B/|J|$ for the
isotropic spin 1/2 dimer were given previously.\cite{ek}  For
$s_1=1/2$ with an arbitrary ${\bm B}$ and $J_j$ for $j=a,b,d,e$,
the rank 4 Hamiltonian matrix is block diagonal, since $s=0,1$ is
a good quantum number.
 The eigenvalues  are given by
\begin{eqnarray}
\epsilon_1&=&-\frac{J_a}{2},\label{epsilon1}\\
\epsilon_n&=&-\frac{J_a}{2}-J+\lambda_n,\qquad
n=2,3,4,\label{epsilon234}
\end{eqnarray}
where
\begin{eqnarray}
0&=&\lambda^3_n+2\lambda^2_nJ_b-\lambda_n[J_d^2-J_b^2+b^2]\nonumber\\
& &-b^2\sin^2\theta[J_b-J_d\cos(2\phi)].\label{halfeigen}
\end{eqnarray}
 For the special cases
  ${\bm B}||\hat{\bm
i}$ for $i=x,y,z$, the $\lambda_n^i$ satisfy
\begin{eqnarray}
\lambda_n^z&=&0,-J_b\pm F_z,\label{lambdaz}\\
\lambda_n^{x,y}&=&-2J_{y,x}, \>\>-J_{x,y}\pm
F_{x,y},\label{lambdaxy}
\end{eqnarray}
where \begin{eqnarray}
 F_i&=&\sqrt{b^2+J_i^2},\label{Fz}\\
 J_{x,y,z}&=&(J_b\pm J_d)/2,\>\> J_d,\label{Jxy}
\end{eqnarray}
respectively, and where $J_x$ ($J_y$) corresponds to the upper
(lower) sign.

 The magnetization for ${\bm B}||\hat{\bm i}$ is given
by
\begin{eqnarray}
M_i&=&\frac{\gamma^2B\sinh(\beta F_i)}{F_i{\cal D}_i},\label{magi} \\
{\cal D}_i&=&\cosh(\beta F_i)+\Delta_i/2,\label{Di}\\
\Delta_{x,y}&=&\exp(-\beta J_{x,y})[\exp(-\beta J)+\exp(2\beta J_{x,y})],\\
\Delta_z&=&\exp(-\beta J_b)[\exp(-\beta J)+1],
\end{eqnarray}
 respectively.
  When the
interactions are written in terms of the less physical $J_f, J_b,
J_d$, and $J_c$, then $J_b\rightarrow J_b+J_f$ and $J_d\rightarrow
J_d+J_c/2$, so that for $s_1=1/2$, $J_f$ and $J_c$ merely
renormalize $J_b$ and $J_d$.  Neither of the single-ion spin
anisotropy terms ${\cal H}_a$ and ${\cal H}_e$ affect the
thermodynamics for $s_1=1/2$. We note that $M_y(J_d)=M_x(-J_d)$
for each $B$, as expected from ${\cal H}_d=-J_d(S_x^2-S_y^2)$.

 From Eqs.
(\ref{epsilon1}), (\ref{epsilon234}), (\ref{lambdaz}), and
(\ref{lambdaxy}),  the single level crossing induction ($s=1$) for
am $s_1=1/2$ dimer  with ${\bm B}||\hat{\bm i}$ occurs at
\begin{eqnarray}
 \gamma  B^{\rm
lc}_{1,1/2}&=&\left\{\begin{array}{cc}\sqrt{J^2+J(J_b\pm
J_d)},&{\bm
B}||\hat{\bm x},\hat{\bm y}\\
\sqrt{(J+J_b)^2-J_d^2},&{\bm B}||\hat{\bm z}\end{array}\right.
,\label{Bstephalf}
\end{eqnarray}
 provided that  $J+J_{x,y}<0$ and $J+J_b<0$,
respectively.

To distinguish the different effects of the  global anisotropy
interactions $J_b$ and $J_d$ that affect the magnetization of
$s_1=1/2$ dimers, in Figs. 2 and 3,  we have respectively plotted
the low-$T$ $ M/\gamma$ versus $\gamma B/|J|$ with $J_b=0.1J,
J_d=0$ and $J_d=0.1J, J_b=0$ for ${\bm B}|||\hat{\bm z}$ (solid)
and ${\bm B}||\hat{\bm x}$ (dashed), along with the isotropic case
$J_b=J_d=0$ (dotted).   From Fig. 2, $J_b<0$ and $J_d=0$ causes a
greater shift to higher $B$ values at the magnetization step with
${\bm B}||\hat{\bm z}$ than for ${\bm B}||\hat{\bm x}$, consistent
with Eq. (\ref{Bstephalf}).  In addition, $J_d$ finite with
$J_b=0$ has a very different effect upon the anisotropy of the
magnetization step, as shown in Fig. 3. Although for ${\bm
B}||\hat{\bm z}$, $B$ at the step is slightly reduced from its
isotropic interaction value, for ${\bm B}||\hat{\bm x}$, the
magnetization step occurs at a larger $B$. These results are
consistent with Eq. (\ref{Bstephalf}).  The midpoint slopes are
universal, in accordance with Eq. (\ref.{slope}).

\begin{figure}
\includegraphics[width=0.45\textwidth]{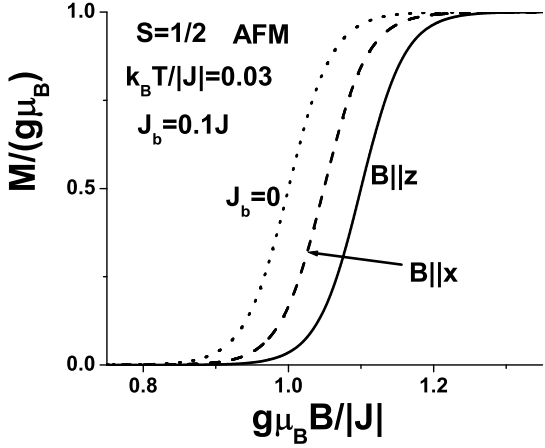}
\caption{Plots of $M/\gamma$ versus $\gamma B/|J|$ at
$k_BT/|J|=0.03$ for the AFM spin 1/2 dimer with $J_b=0.1J$,
$J_d=0$, with ${\bm B}||\hat{\bm z}$ (solid), ${\bm B}||\hat{\bm
x}$ (dashed), along with the isotropic case $J_b=J_d=0$ (dotted).}
\label{fig2}
\end{figure}

\begin{figure}
\includegraphics[width=0.45\textwidth]{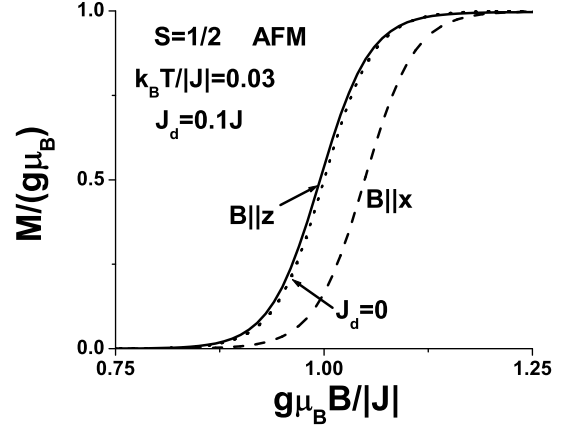}
\caption{Plots of $M/\gamma$ versus $\gamma B/|J|$ at
$k_BT/|J|=0.03$ for the AFM spin 1/2 dimer with $J_d=0.1J$,
$J_b=0$, with ${\bm B}||\hat{\bm z}$ (solid), ${\bm B}||\hat{\bm
x}$ (dashed), along with the isotropic case $J_b=J_d=0$ (dotted).}
\label{fig3}
\end{figure}

The specific heat of an $s_1=1/2$ dimer with ${\bm B}||\hat{\bm
i}$ is
\begin{eqnarray}
C_{Vi}&=&\frac{k_B\beta^2{\cal N}_i}{{\cal D}_i^2},\end{eqnarray}
 where the ${\cal D}_i$ are given by Eq. (\ref{Di}), and
 the ${\cal N}_i$ are given in Appendix A.
Plots at
 low $T$ of $C_V/k_B$ versus $\gamma B/|J|$ for $s_1=1/2$ dimers
with the corresponding global anisotropies $J_b=0.1J$ and
$J_d=0.1J$ are shown in Figs. 4 and 5, respectively.  We note the
universal curve shapes, but non-universal level-crossing
positions, in quantitative agreement with Eqs.
(\ref{CVzeroes})-(\ref{peakheight}). In Fig. 4, the positions of
the maxima and the central minimum in $C_V$ track that of the
magnetization step in Fig. 2 with the same parameters. With
$J_d=0.1J$, the behaviors in $C_V$ and $M$ for ${\bm B}||\hat{\bm
x}$ are also very similar. However, there is a slight difference
in the behaviors for ${\bm B}||\hat{\bm z}$. Note that $M$ (Fig.
3) shows a slight reduction for ${\bm B}||\hat{\bm z}$ in the
induction required for the step, whereas $C_V$ (Fig. 5) shows a
slight increase in the positions of the peaks.  This detail  only
appears when $J_d\ne0$, for which the effective temperature is
slightly higher than for $J_d=0$.

\begin{figure}
\includegraphics[width=0.45\textwidth]{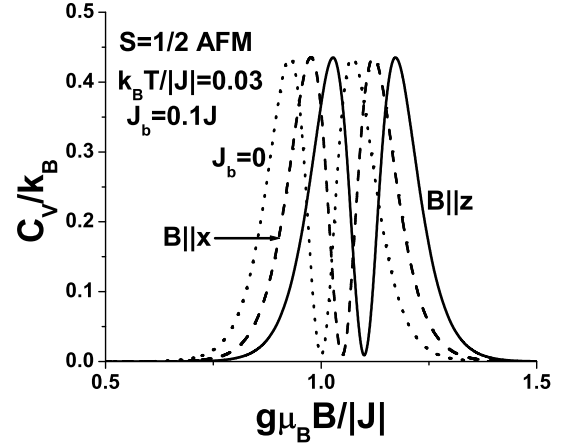}
\caption{Plot of $C_V/k_B$ versus $\gamma B/|J|$ for the AFM spin
1/2 dimer with $J_b=0.1J, J_d=0$ at $k_BT/|J|=0.03$ with the same
curve notation as in Fig. 2.}\label{fig4}
\end{figure}

\begin{figure}
\includegraphics[width=0.45\textwidth]{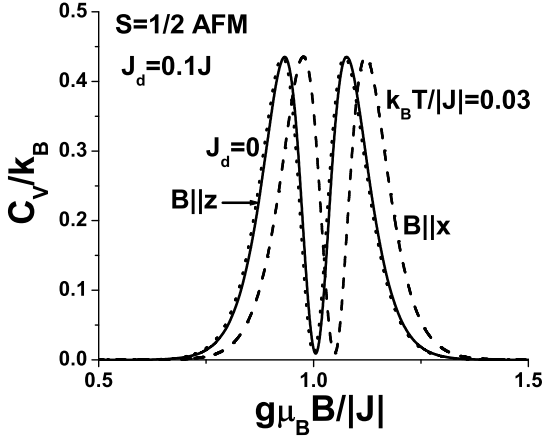}
\caption{Plot of $C_V/k_B$ versus $\gamma B/|J|$ for the AFM spin
1/2 dimer with $J_d=0.1J$, $J_b=0$  at $k_BT/|J|=0.03$ with the
same curve notation as in Fig. 3.}\label{fig5}
\end{figure}

\section{Analytic and numerical results for spin 1}

For dimers with $s_1=1$, the allowed $s$ values are $s=0,1,2$. The
three $s=1$ states are decoupled from the six remaining $s=0,2$
states. They satisfy a cubic equation given in Appendix A. For
${\bm B}||\hat{\bm i}$ with $i=x,y,z$, this cubic equation
simplifies to a linear and a quadratic equation, as for the $s=1$
eigenstates of $s_1=1/2$ dimers.

The remaining six eigenstates corresponding nominally to $s=0,2$
are in general all mixed.  The  matrix  leading to the hexatic
equation from which the six eigenvalues can be obtained is given
in Appendix A. That is sufficient to evaluate the eigenvalues for
$s_1=1$ using symbolic manipulation software. When combined with
the three $s=1$ eigenvalues, one can then use Eqs. (\ref{sh}) and
(\ref{mag}) to obtain the resulting exact magnetization and
specific heat at an arbitrary ${\bm B}$. The combined nine
eigenvalues depend upon all four anisotropy parameters $J_j$ for
$j=a,b,d,e$.

 To the extent that the eigenvalues can be obtained from the
solutions to either linear or quadratic equations, the expressions
for the $B_{s,1}^{\rm lc}$ are simple, and are given in Appendix
A. In the global anisotropy case $J_a=J_e=0$, the first level
crossing induction $B_{1,1}^{\rm lc}$ with ${\bm B}||\hat{\bm i}$
is identical to $B_{1,1/2}^{\rm lc}$, the level crossing with
$s_1=1/2$ given by Eq. (\ref{Bstephalf}). For comparison, we
expand the first and second level crossing inductions to first
order in each of the $J_j$ for $j=a,b,d,e$ for ${\bm B}||\hat{\bm
i}$ where $i=x,y,z$,
\begin{eqnarray}
\gamma B_{1,1,z}^{{\rm
lc}(1)}&=&-J+\frac{J_a}{3}-J_b,\label{b11z}\\
\gamma B_{1,1,x,y}^{{\rm lc}
(1)}&=&-J-\frac{J_a}{6}-\frac{J_b}{2}\mp\frac{1}{2}(J_d-J_e),\label{b11xy}\\
\gamma B_{2,1,z}^{{\rm lc} (1)}&=&-2J-J_a-3J_b,\label{b21z}\\
\gamma B_{2,1,x,y}^{{\rm lc}
(1)}&=&-2J+\frac{J_a}{2}-\frac{J_b}{2}\mp\frac{5J_d}{2}\mp\frac{3J_e}{2},\label{b21xy}
\end{eqnarray}
where the upper (lower) sign is for ${\bm B}||\hat{\bm x}$ (${\bm
B}||\hat{\bm y}$), respectively.

From these simple first-order results, it is possible to
understand the qualitatively different  behavior obtained with
local, single-ion, anisotropy from that obtained with global
anisotropy. In the isotropic case $J_j=0\>\>\forall j$, the first
and second level crossings occur at $-J$ and $-2J$, respectively.
For each induction direction,  the signs of the $J_b$ and $J_d$
contributions to $B_{2,1}^{{\rm lc}(1)}+2J$ and $B_{1,1}^{{\rm
lc}(1)}+J$ are the same, whereas the signs of the $J_a$ and $J_e$
contributions to  $B_{2,1}^{{\rm lc}(1)}+2J$ and $B_{1,1}^{{\rm
lc}(1)}+J$ are the {\it opposite}.

In Figs. 6-15, we plot $M/\gamma$  and $C_V/k_B$ versus $\gamma
B/|J|$ for
  five low-$T$ cases of AFM $s_1=1$ dimers, taking
 $k_BT/|J|=0.03$.   The $M/\gamma$ curves all exhibit the universal features
 predicted by Eqs. (\ref{Msteps}) and (\ref{slope}).
 The corresponding $C_V/k_B$  curves
 also obey the universal features predicted in Eqs. (\ref{CVzeroes})-(\ref{peakheight}). In these figures, only one of the five
 anisotropy interactions $J_j$ is non-vanishing, and we take
 $J_j/J=0.1$, for $j= b, d, a, e,$ and $c$, respectively.  Unlike the case of $s_1=1/2$
 dimers,  for which $J_c$ merely renormalizes $J_d$, for $s_1=1$
 all of these interactions lead to distinct anisotropy
 effects in the low-$T$ magnetization and specific heat.  Of
 course, since the case
 $J_c=0.1J$ (and the remaining $J_j=0$) can be evaluated by setting
  $J_d=0.05J$ and $J_e=-0.05J$ for any $s_1$,  it is not really distinct from the other anisotropy interactions,
 but only represents that particular combination of global and
 local azimuthal anisotropy interactions.

 \begin{figure}
\includegraphics[width=0.45\textwidth]{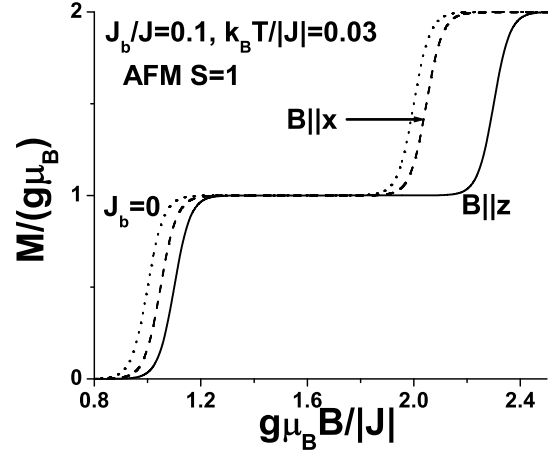}
\caption{Plot at $k_BT/|J|=0.03$ and $J_b/J=0.1$ of $M/\gamma$
versus $\gamma B/|J|$ for the AFM spin 1 dimer. Curves for ${\bm
B}||\hat{\bm z}$ (solid),  ${\bm B}||\hat{\bm x}$ (dashed), and
the isotropic case ($J_b=0$, dotted) are shown.}\label{fig6}
\end{figure}

 \begin{figure}
\includegraphics[width=0.45\textwidth]{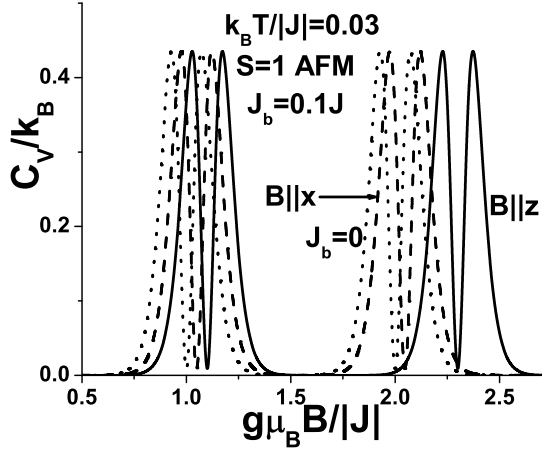}
\caption{Plot at $k_BT/|J|=0.03$ and $J_b/J=0.1$ of $C_V/k_B$
versus $\gamma B/|J|$ for the AFM spin 1 dimer. Curves for ${\bm
B}||\hat{\bm z}$ (solid),  ${\bm B}||\hat{\bm x}$ (dashed), and
the isotropic case ($J_b=0$, dotted) are shown.}\label{fig7}
\end{figure}

 We first show the results for the global anisotropy interactions.
  In Figs. 6 and 7, we plot $M(B)$ and $C_V(B)$ for $J_b=0.1J$ (and all other $J_j=0$) at
  $k_BT/|J|=0.03$ for the AFM $s_1=1$ dimer.  The
  level-crossing
  induction increases monotonically
  with level-crossing number, with the largest induction required occurring for
  ${\bm B}||\hat{\bm z}$, as for $s_1=1/2$, quantitatively consistent with Eqs. (\ref{b11z})-(\ref{b21xy}).
    In Figs. 8 and 9, the
  corresponding results for $J_d=0.1J$ (and the other $J_j=0$) are
  shown.  Again, the level-crossing induction
  increases monotonically with level-crossing number,
  and the largest
  induction
  required for each level crossing occurs for ${\bm B}||\hat{\bm
  x}$, quantitatively consistent with Eqs. (\ref{b11z})-(\ref{b21xy}), as for $s_1=1/2$.
  We note that the solid curves for ${\bm
  B}||\hat{\bm z}$ are nearly identical with the dotted curves for
  the isotropic case $J_j=0\forall j$.  However, there is a very
  small distinction between these $M$ and $C_V$ curves.  This
  distinction is also present for $s_1=1/2$, as shown in Figs. 3
  and 5, and is a detail of the shapes of the $M$ and $C_V$ curves
  arising from a slightly higher effective temperature with
  $J_d\ne0$ than in the other cases studied.

\begin{figure}
\includegraphics[width=0.45\textwidth]{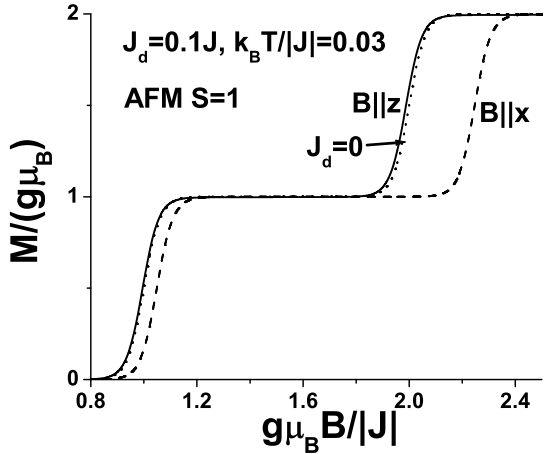}
\caption{Plot at $k_BT/|J|=0.03$ and $J_d/J=0.1$ of ${\bm
M}/\gamma$ versus $\gamma B/|J|$ for the AFM spin 1 dimer. Curves
for ${\bm B}||\hat{\bm z}$ (solid),  ${\bm B}||\hat{\bm x}$
(dashed), and the isotropic case ($J_d=0$, dotted) are
shown.}\label{fig8}
\end{figure}

 \begin{figure}
\includegraphics[width=0.45\textwidth]{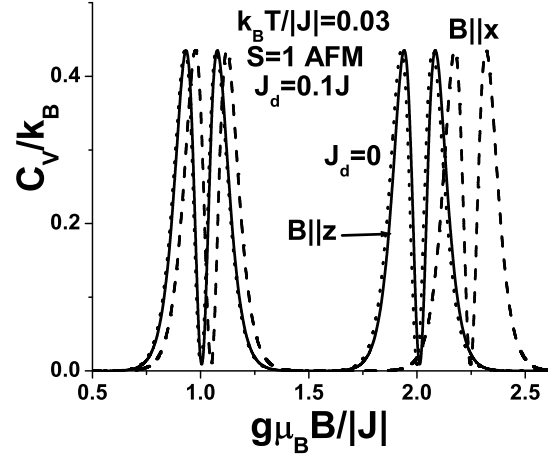}
\caption{Plot at $k_BT/|J|=0.03$ and $J_d/J=0.1$ of $C_V/k_B$
versus $\gamma B/|J|$ for the AFM spin 1 dimer. Curves for ${\bm
B}||\hat{\bm z}$ (solid),  ${\bm B}||\hat{\bm x}$ (dashed), and
the isotropic case ($J_b=0$, dotted) are shown.}\label{fig9}
\end{figure}

  Next, in Figs. 10-13, we show the corresponding curves for
  $M(B)$ and $C_V(B)$ at low $T$ for the local anisotropies
  $J_a=0.1J$ and $J_e=0.1J$, respectively, with the other $J_j=0$.  In contrast
  to the global anisotropies, the increase in $B$ required with the level
  crossing  is not monotonic in the level-crossing number,  quantitatively
  consistent with Eqs. (\ref{b11z})-(\ref{b21xy}).  In particular,
  we note that in each of these figures, the deviations in the level crossings
  $B_{2,1}^{\rm lc}+2J$ and $B_{1,1}^{\rm
lc}+J$ are   opposite in sign.

  A careful examination of the
  exact formula for $C_V$ reveals that it vanishes at
  precisely the level crossing inductions as $T\rightarrow0$.  Similarly,  the
  magnetization curve
  attains its midpoint values $\frac{1}{2}$ and $\frac{3}{2}$ between the two neighboring steps at
  the level crossing inductions
  as $T\rightarrow0$.

  \begin{figure}
\includegraphics[width=0.45\textwidth]{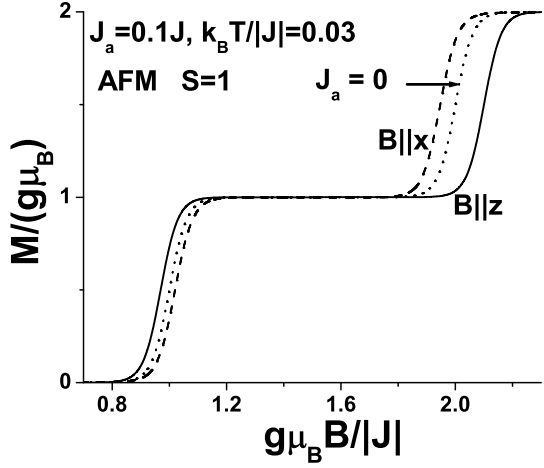}
\caption{Plot at $k_BT/|J|=0.03$ and $J_a/J=0.1$ of $M/\gamma$
versus $\gamma B/|J|$ for the AFM spin 1 dimer. Curves for ${\bm
B}||\hat{\bm z}$ (solid),  ${\bm B}||\hat{\bm x}$ (dashed), and
the isotropic case ($J_a=0$, dotted) are shown.}\label{fig10}
\end{figure}

 \begin{figure}
\includegraphics[width=0.45\textwidth]{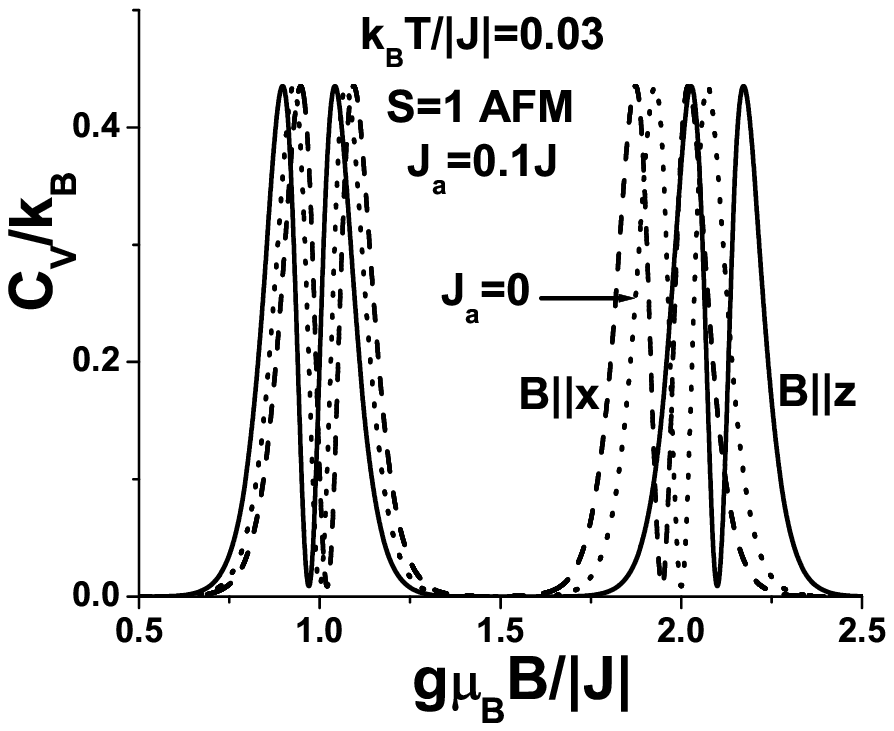}
\caption{Plot at $k_BT/|J|=0.03$ and $J_a/J=0.1$ of $C_V/k_B$
versus $\gamma B/|J|$ for the AFM spin 1 dimer. Curves for ${\bm
B}||\hat{\bm z}$ (solid),  ${\bm B}||\hat{\bm x}$ (dashed), and
the isotropic case ($J_b=0$, dotted) are shown.}\label{fig11}
\end{figure}

\begin{figure}
\includegraphics[width=0.45\textwidth]{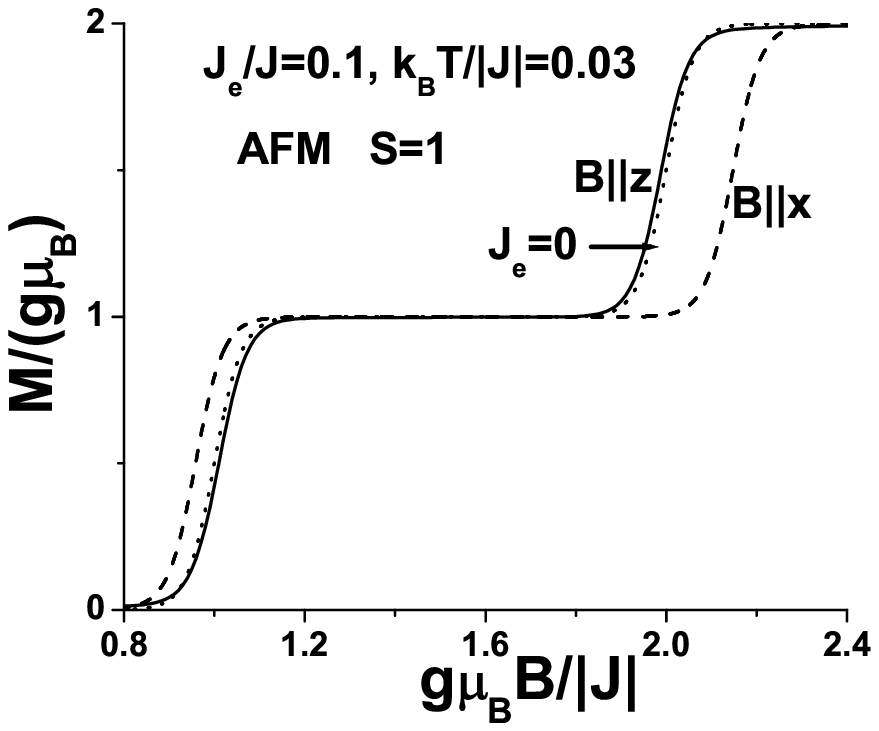}
\caption{Plot at $k_BT/|J|=0.03$ and $J_e/J=0.1$ of $M/\gamma$
versus $\gamma B/|J|$ for the AFM spin 1 dimer. Curves for ${\bm
B}||\hat{\bm z}$ (solid),  ${\bm B}||\hat{\bm x}$ (dashed), and
the isotropic case ($J_e=0$, dotted) are shown.}\label{fig12}
\end{figure}

 \begin{figure}
\includegraphics[width=0.45\textwidth]{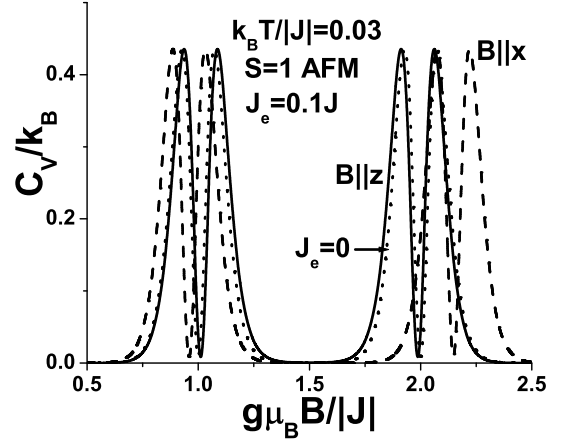}
\caption{Plot at $k_BT/|J|=0.03$ and $J_b/J=0.1$ of $C_V/k_B$
versus $\gamma B/|J|$ for the AFM spin 1 dimer. Curves for ${\bm
B}||\hat{\bm z}$ (solid),  ${\bm B}||\hat{\bm x}$ (dashed), and
the isotropic case ($J_b=0$, dotted) are shown.}\label{fig13}
\end{figure}

  Finally, in Figs. 14 and 15, we show the corresponding $M(B)$  and $C_V(B)$ curves for $J_c=0.1J$,
  corresponding to the parameter
  choices $J_e=-0.05J$ and $J_d=0.05J$.  With this combination,
  the magnetization  and specific heat curves shown in Figs. 14
  and 15
  resemble those for $J_d=0.1J$ shown in Figs. 8 and 9, except that
  the second magnetization steps  and the second set of specific heat peaks have nearly the same behavior as do the first
  magnetization steps and specific heat double peaks.

  In short, the case $s_1=1$, for which the exact expressions from
  which the eigenvalues can be readily obtained are sufficiently
  short as to be writable on paper, are sufficient to exhibit the
  very different behaviors obtained from the single-ion, local
  spin anisotropy interactions from those obtained from the global
  spin anisotropy interactions.  As $s_1$ increases
  beyond 1, the situation becomes not only more complicated, but also
   more interesting, as shown
  in the following.

\begin{figure}
\includegraphics[width=0.45\textwidth]{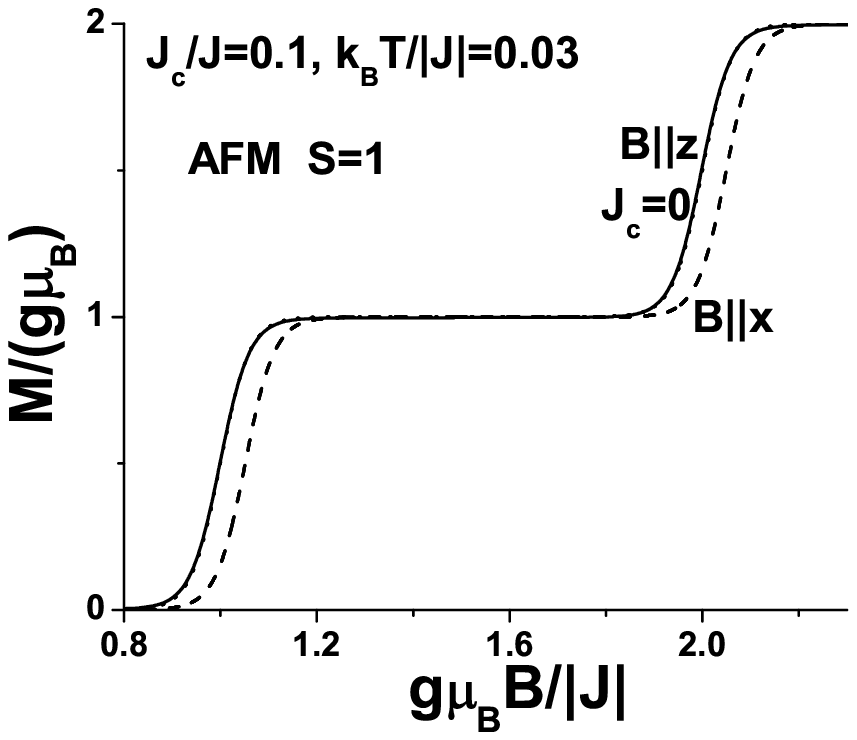}
\caption{Plot at $k_BT/|J|=0.03$ and $J_c/J=0.1$ of $M/\gamma$
versus $\gamma B/|J|$ for the AFM spin 1 dimer. Curves for ${\bm
B}||\hat{\bm z}$ (solid),  ${\bm B}||\hat{\bm x}$ (dashed), and
the isotropic case ($J_c=0$, dotted) are shown.}\label{fig14}
\end{figure}

 \begin{figure}
\includegraphics[width=0.45\textwidth]{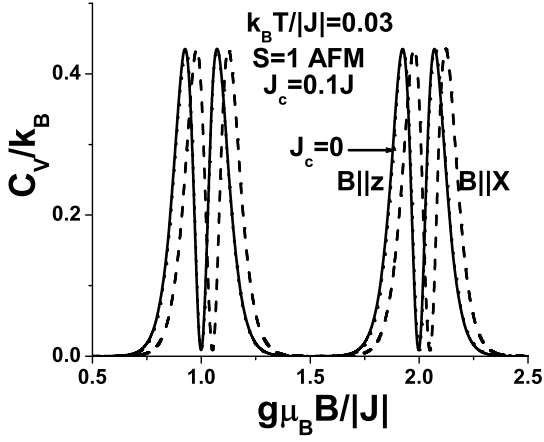}
\caption{Plot at $k_BT/|J|=0.03$ and $J_c/J=0.1$ of $C_V/k_B$
versus $\gamma B/|J|$ for the AFM spin 1 dimer. Curves for ${\bm
B}||\hat{\bm z}$ (solid),  ${\bm B}||\hat{\bm x}$ (dashed), and
the isotropic case ($J_b=0$, dotted) are shown.}\label{fig15}
\end{figure}

\section{Exact Numerical Results for spin 5/2}

For $s_1=5/2$, one of the cases of greatest experimental interest,
when ${\cal H}_a$ and ${\cal H}_e$ are present, none of the
allowed $s, m$ values is a true quantum number.  That is, ${\cal
H}_a$ and ${\cal H}_e$ cause all of the states with nominally odd
or even $s$  to mix with one another. For ${\bm B}||\hat{\bm i}$
for $i=x, y, z$, this simplifies in the  crystal representation as
for $s_1=1$, since only states with odd or even $m$ in the
appropriately chosen representation can mix. By using symbolic
manipulation software, it is possible to solve for the exact
eigenvalues of the $s_1=5/2$ dimer.  However, because the analytic
expressions for the eigenvalues are much more complicated than
those for $s_1=1$ presented in Appendix A, we shall not attempt to
present them, but will instead focus upon their numerical
evaluation for specific cases.

To first order in the $J_j$, the first three level crossings for
${\bm B}||\hat{\bm i}$ with $i=x, y, z$ are
\begin{eqnarray}
\gamma B_{1,5/2,z}^{{\rm lc}(1)}&=&-J+\frac{32 J_a}{15}-J_b,\label{B52step1}\\
\gamma B_{1,5/2,x,y}^{{\rm
lc}(1)}&=&-J-\frac{16J_a}{15}-\frac{J_b}{2}\mp\frac{J_d}{2}
\pm\frac{16J_e}{5},\\
\gamma B_{2,5/2,z}^{{\rm lc}(1)}&=&-2J-\frac{8J_a}{35}-3J_b,\\
\gamma B_{2,5/2,x,y}^{{\rm
lc}(1)}&=&-2J+\frac{4J_a}{35}-\frac{J_b}{2}\mp\frac{5J_d}{2}\mp\frac{12J_e}{35},\\
\gamma B_{3,5/2,z}^{{\rm lc}(1)}&=&-3J-\frac{106J_a}{63}-5J_b,\\
\gamma B_{3,5/2,x,y}^{{\rm
lc}(1)}&=&-3J+\frac{53J_a}{63}-\frac{J_b}{2}\mp\frac{9J_d}{2}\mp\frac{53J_e}{21}.\label{B52step3}
\end{eqnarray}

\begin{figure}
\includegraphics[width=0.45\textwidth]{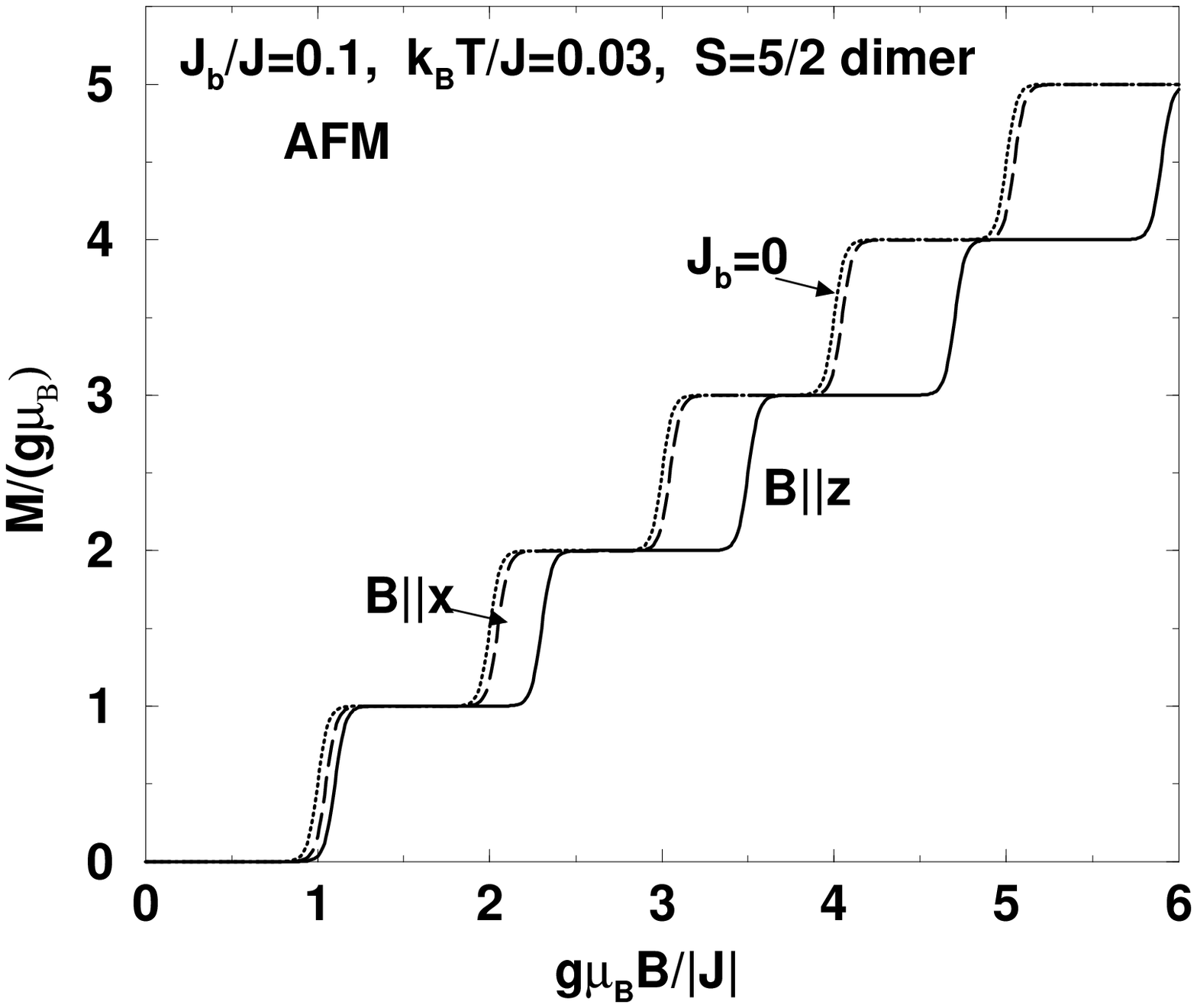}
\caption{Plot  at $k_BT/|J|=0.03$ of $M/\gamma$ versus $\gamma
B/|J|$ for the AFM spin 5/2 dimer with $J_b/J=0.1$. ${\bm
B}||\hat{\bm z}$ (solid), ${\bm B}||\hat{\bm x}$ (dashed), and the
isotropic case ($J_b=0$, dotted) are shown.}\label{fig16}
\end{figure}

 \begin{figure}
\includegraphics[width=0.45\textwidth]{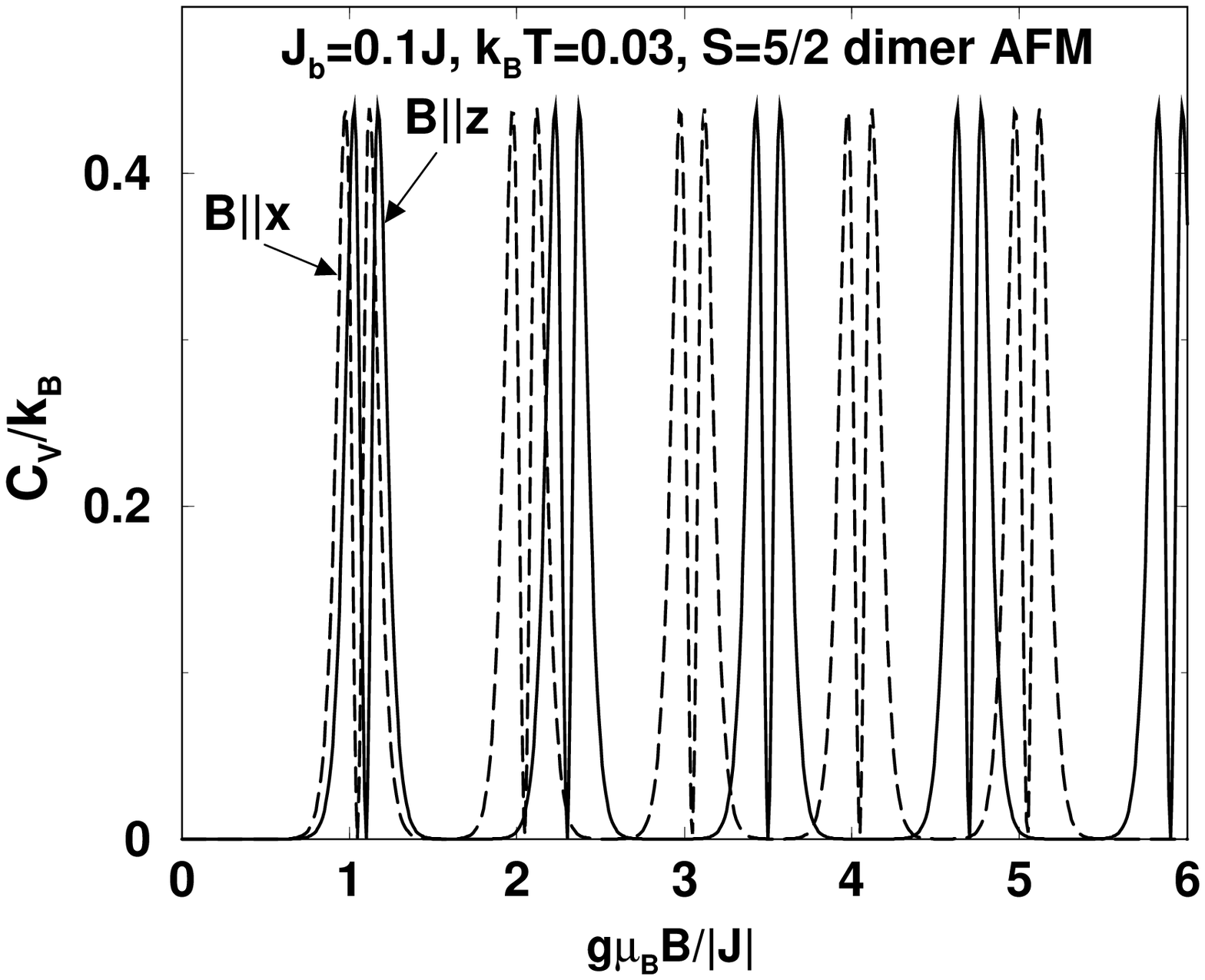}
\caption{Plot at $k_BT/|J|=0.03$ and $J_b/J=0.1$ of $C_V/k_B$
versus $\gamma B/|J|$ for the AFM spin 5/2 dimer. Curves for ${\bm
B}||\hat{\bm z}$ (solid),  ${\bm B}||\hat{\bm x}$ (dashed), and
the isotropic case ($J_b=0$, dotted) are shown.}\label{fig17}
\end{figure}

\begin{figure}
\includegraphics[width=0.45\textwidth]{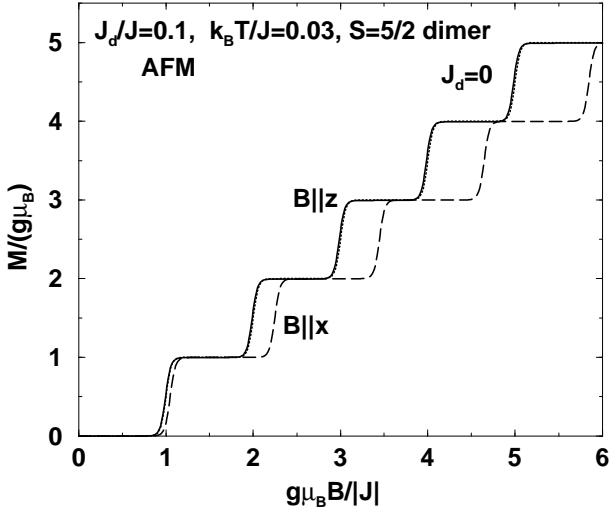}
\caption{Plot  of $M/\gamma$ versus $\gamma B/|J|$ for the AFM
spin 5/2 dimer at $k_BT/|J|=0.03$ with $J_d/J=0.1$. Curves for
${\bm B}||\hat{\bm z}$ (solid),  ${\bm B}||\hat{\bm x}$ (dashed),
and the isotropic case ($J_b=0$, dotted) are shown.}\label{fig18}
\end{figure}

 \begin{figure}
\includegraphics[width=0.45\textwidth]{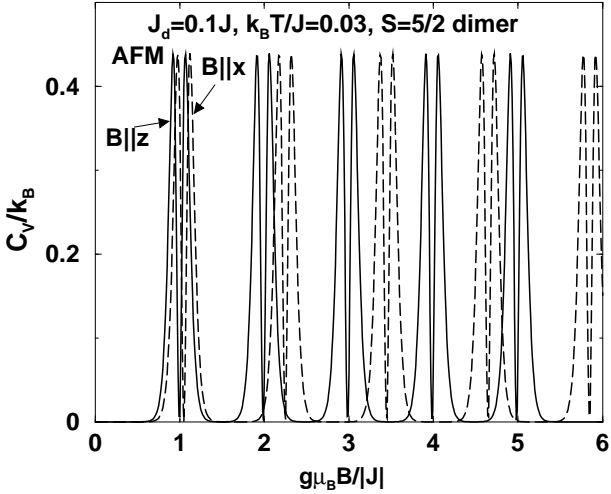}
\caption{Plot at $k_BT/|J|=0.03$ and $J_d/J=0.1$ of $C_V/k_B$
versus $\gamma B/|J|$ at $k_BT/|J|=0.03$ for the AFM spin 5/2
dimer. Curves for ${\bm B}||\hat{\bm z}$ (solid),  ${\bm
B}||\hat{\bm x}$ (dashed), and the isotropic case ($J_b=0$,
dotted) are shown.}\label{fig19}
\end{figure}

In Figs. 16-23, we plot $M/\gamma$  and $C_V/k_B$ versus $\gamma
B/|J|$ for four low-$T$ cases of AFM $s_1=5/2$ dimers, $J_b=0.1J$,
$J_d=0.1J$, $J_a=0.1J$, and $J_c=0.1J$, respectively, and the
other $J_j=0$, taking $k_BT/|J|=0.03$.  In Fig. 24, examples of
$M(\theta)$ at fixed $B$ and $\phi=0$ are shown. Figures 16-24 are
sufficient to distinguish the more interesting local spin
anisotropy effects in AFM dimers with higher $s_1$ values from the
non-existent or less interesting ones present  with $s_1=1/2,1$,
respectively. In Figs. 16, 18, 20, and 22, the solid and dashed
curves represent the cases of ${\bm B}||\hat{\bm z}$ and ${\bm
B}||\hat{\bm x}$, and the dotted curve is the isotropic case,
$J_j=0\forall j$, as in Figs. 2-15. Because of the number of peaks
in the specific heat curves, in Figs. 17, 19, 21, and 23, we
ol=nly showed $C_V/k_B$ versus $\gamma B/|J|$ for ${\bm
B}||\hat{\bm z}$ (solid) and ${\bm B}||\hat{\bm x}$ (dashed).  The
isotropic curve with $J_j=0\>\>L\forall j$ was published
previously.\cite{ek} That curve has minima at $\gamma B/|J|=s$ for
$s=1,\ldots,5$, each of which is central to double peaks. ${\cal
H}_a$ and ${\cal H}_b$ are invariant under $x\leftrightarrow y$,
so curves for ${\bm B}||\hat{\bm x}$ in Figs. 16, 17, 20, and 21
are identical to those for ${\bm B}||\hat{\bm y}$. Since ${\cal
H}_d$ and ${\cal H}_c$ are odd under $x\leftrightarrow y$, the
${\bm B}||\hat{\bm x}$ curves in Figs. 18, 19, 22, and 23
correspond to ${\bm B}||\hat{\bm y}$ with $J_d/J=-0.1J$,
$J_c/J=-0.1$, respectively.

 We first examine the global spin anisotropy
effects of ${\cal H}_b$ and ${\cal H}_d$ in Figs. 16-19.
 These
figures exhibit the same behavior shown for $s_1=1/2,1$ in the
corresponding Figs. 2-9.  Note that  $B_{s,5/2}^{\rm
lc}(\theta,0)$ is largest for $\theta=\pi/2$ with $J_b=0.1J$ and
for $\theta=0$ with $J_d=0.1J$, and increases monotonically with
$s$, as for $s_1=1$ in both cases, nearly quantitatively
consistent with Eqs. (\ref{B52step1})-(\ref{B52step3}). By
contrast, $B_{s,52}^{\rm lc}(\theta,0)$ for $\theta=\pi/2$ with
$J_b=0.1J$ and for $\theta=0$ for $J_d=0.1J$ are nearly
indistinguishable from the isotropic case, also nearly
quantitatively consistent with Eqs.
(\ref{B52step1})-(\ref{B52step3}).

In contrast, the local field anisotropy interactions show very
different and much more interesting behaviors.  In Figs. 20-21, we
present our results for the effects of the single-ion axial
anisotropy interaction ${\cal H}_a$, Eq. (\ref{lz}). As in Fig. 10
for $s_1=1$, $B_{s,5/2}^{\rm lc}(\theta,0)+sJ$ changes sign with
increasing $s$.  $B_{1,5/2}^{\rm lc}(0,0)+J<0$, whereas for
$s\ge2$, $B_{s,5/2}^{\rm lc}(0,0)+sJ>0$ and increases
monotonically with  $s$.   For $\theta=\pi/2$, nearly the opposite
situation occurs. $B_{s,5/2}^{\rm lc}(\pi/2,0)+sJ$ is positive for
$s=1$ and decreases monotonically with increasing $s$.  In both
cases,$|B_{s,5/2}^{\rm lc}(\theta,0)+sJ|$ is a minimum for $s=2$.
 These local axial anisotropy effects,
consistent with Eqs. (\ref{B52step1})-(\ref{B52step3}), are very
different than the global anisotropy ones pictured in Figs. 16-19.
They are also much richer and interesting than the corresponding
case for $s_1=1$ pictured in Figs. 10 and 11.

In Figs. 22-23, we  present our $M/\gamma$ versus $\gamma B/|J|$
results for the case of the local azimuthally anisotropic exchange
interaction, ${\cal H}_c$, Eq. (\ref{Hc}), evaluated for AFM
dimers at $k_BT/|J|=0.03$ with $J_c=0.1J$ and the remaining
$J_j=0$.  $B_{s,5/2}^{{\rm lc}(1)}$ for this case is obtained from
Eqs. (\ref{B52step1})-(\ref{B52step3}) by setting $J_d, J_e=\pm
J_c/2$, respectively, and $J_a=J_b=0$.  $B_{s,5/2}^{\rm lc}(0,0)$
is nearly indistinguishable from the isotropic case, as in Figs.
12 and 13 for $s_1=1$, except for some minor curve shape effects
far from the step midpoints.  $B_{s,5/2}^{\rm lc}(\pi/2,0)+sJ$ is
always positive, as for $s_1=1$ shown in Figs. 12 and 13, but has
a minimum at $s=3$.   This is also in stark contrast to the
monotonic global anisotropy behavior for $s_1=5/2$ seen in Figs.
16-19. Both of these behaviors are nearly quantitatively
consistent with Eqs. (\ref{B52step1})-(\ref{B52step3}) as modified
to include $J_c$.

\begin{figure}
\includegraphics[width=0.45\textwidth]{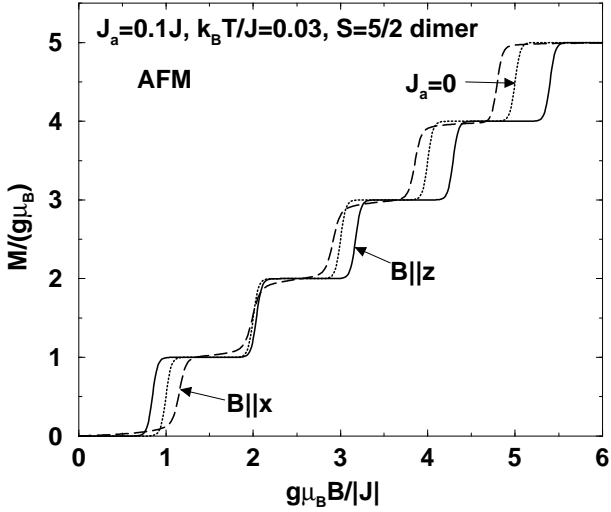}
\caption{Plot at $k_BT/|J|=0.03$ and $J_a/J=0.1$ of $M/\gamma$
versus $\gamma B/|J|$ for the AFM spin 5/2 dimer. Curves for ${\bm
B}||\hat{\bm z}$ (solid),  ${\bm B}||\hat{\bm x}$ (dashed), and
the isotropic case ($J_a=0$, dotted) are shown.}\label{fig20}
\end{figure}

 \begin{figure}
\includegraphics[width=0.45\textwidth]{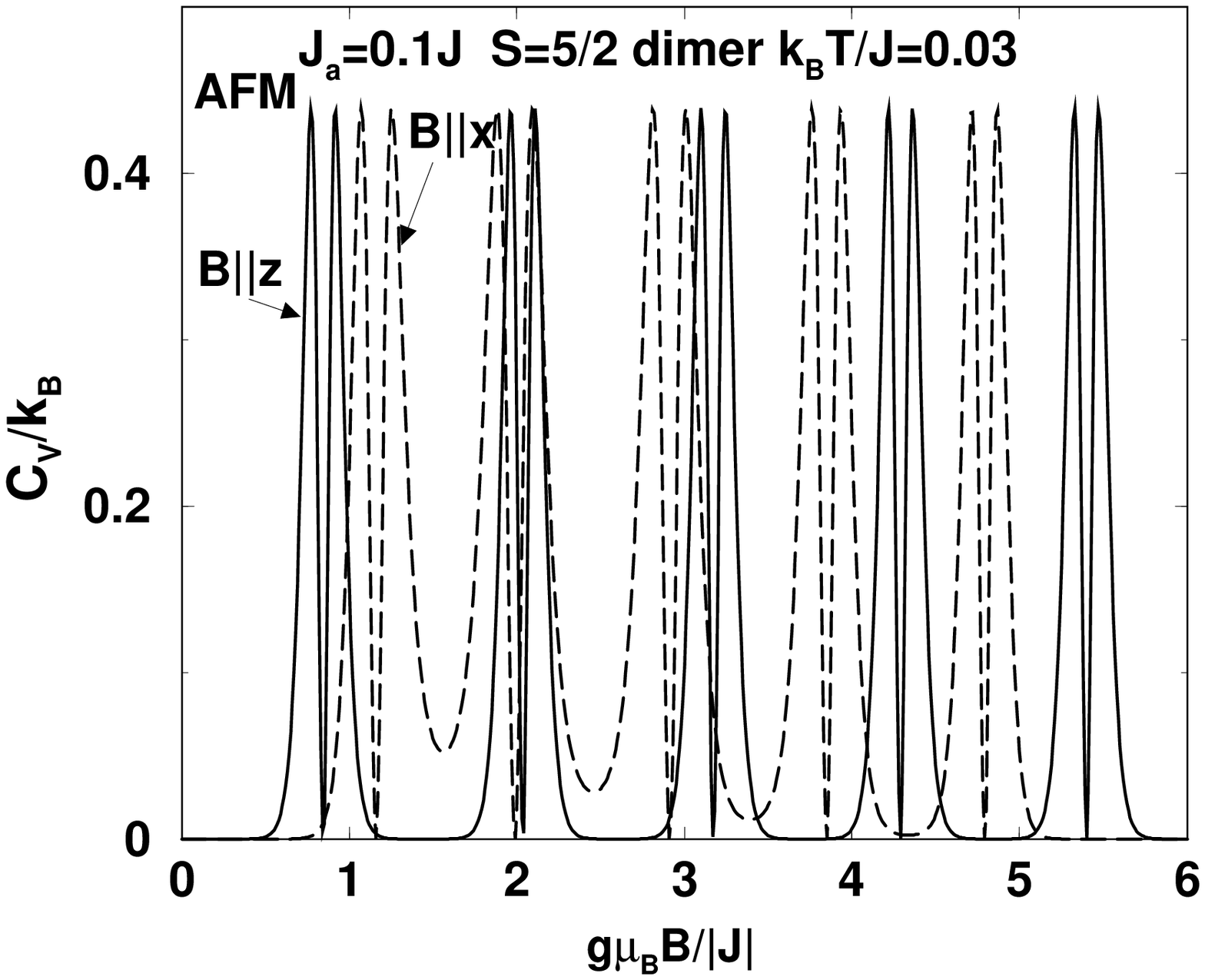}
\caption{Plot at $k_BT/|J|=0.03$ and $J_a/J=0.1$ of $C_V/k_B$
versus $\gamma B/|J|$ at $k_BT/|J|=0.03$ for the AFM spin 5/2
dimer. Curves for ${\bm B}||\hat{\bm z}$ (solid),  ${\bm
B}||\hat{\bm x}$ (dashed), and the isotropic case ($J_b=0$,
dotted) are shown.}\label{fig21}
\end{figure}

\begin{figure}
\includegraphics[width=0.45\textwidth]{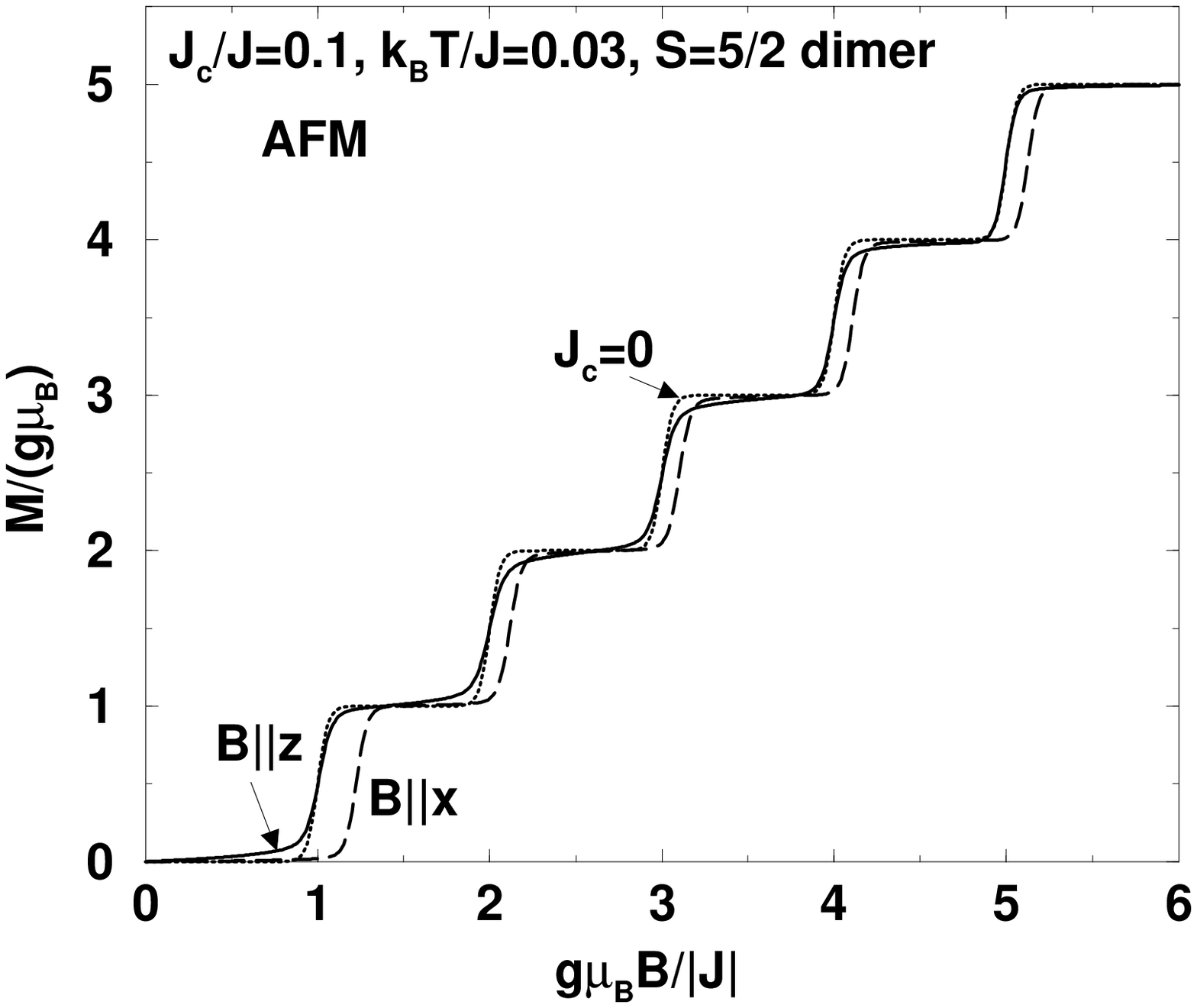}
\caption{Plot  at $k_BT/|J|=0.03$ of $M/\gamma$ versus $\gamma
B/|J|$ for the AFM spin 5/2 dimer with $J_c/J=0.1$. ${\bm
B}||\hat{\bm z}$ (solid), ${\bm B}||\hat{\bm x}$ (dashed), and the
isotropic case ($J_b=0$, dotted) are shown.}\label{fig22}
\end{figure}

 \begin{figure}
\includegraphics[width=0.45\textwidth]{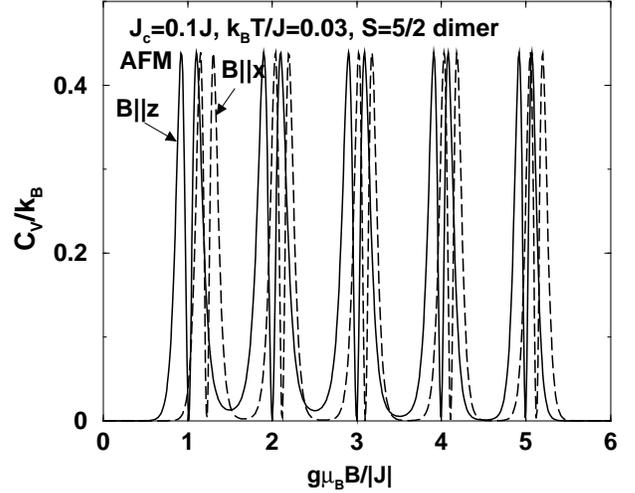}
\caption{Plot at $k_BT/|J|=0.03$ and $J_c/J=0.1$ of $C_V/k_B$
versus $\gamma B/|J|$ at $k_BT/|J|=0.03$ for the AFM spin 5/2
dimer. Curves for ${\bm B}||\hat{\bm z}$ (solid),  ${\bm
B}||\hat{\bm x}$ (dashed), and the isotropic case ($J_b=0$,
dotted) are shown.}\label{fig23}
\end{figure}

In contrast, the local field anisotropy interactions show very
different behaviors.  In Figs. 20 and 21, our results for the
effects of the single-ion axial anisotropy interaction ${\cal
H}_a$, Eq. (\ref{lz}), are shown.  As  in Figs. 10 and 11, the
induction anisotropy effects of ${\cal H}_a$ change sign with
increasing $B$. In Fig. 20, for ${\bm B}||\hat{\bm z}$ the first
magnetization step appears at a lower value of $|{\bm B}|$ than in
the isotropic case, and for steps 3-5, there is a monotonic
increase in the extra field required for each step. The opposite
is true for ${\bm B}||\hat{\bm x}$, for which the first step
appears at a larger $|{\bm B}|$ value than for the isotropic case,
and subsequent magnetization steps appear at monotonically
decreasing values of $|{\bm B}|$. A cross-over occurs at about the
second step, for which the effects of this type of anisotropy are
small.  The corresponding shifts in the positions of the double
peaks in the specific heat are shown in Fig. 21. These local axial
anisotropy effects are very different than the global anisotropy
ones pictured in Figs. 16-19.

In Figs. 22 and 23, we  present our ${\bm M}$ and $C_V$ results
for the case of the local azimuthally anisotropic exchange
interaction, ${\cal H}_c$, Eq. (\ref{Hc}), evaluated for AFM
dimers at $k_BT/|J|=0.03$ with $J_c=0.1J$, or equivalently with
$J_d=0.05J$ and $J_e=-0.05J$ (and the remaining $J_j=0$). For
${\bm B}||\hat{\bm z}$, there is almost no change  from the
isotropic case, as occurred with the global azimuthal anisotropy
pictured in Figs. 128 and 19. For ${\bm B}||\hat{\bm x}$, the
field required for each step is larger than in the isotropic
interaction case, as in Figs. 14 and 15  for $s_1=1$, and for
global azimuthal anisotropy shown in Figs. 18 and 19. However, in
this case, the extra induction required for each level crossing is
non-monotonic in the crossing number, with the largest extra
induction required for the first  crossing, and the minimum extra
induction required for the intermediate, third  crossing. This is
in stark contrast to the monotonic   global anisotropy behavior
for $s_1=5/2$ seen in Figs. 16-19.  Although not pictured
explicitly, the behavior for $J_e=0.1J$ with the other $J_j=0$ is
rather like that of the $J_a=0.1J$ curves pictured in Figs. 20 and
21 with $\hat{\bm x}\leftrightarrow\hat{\bm z}$, differing in ways
similar to those differences between the $J_a=0.1J$ and $J_e=0.1J$
curves pictured for $s_1=1$ in Figs. 10-13.

\begin{figure}
\includegraphics[width=0.45\textwidth]{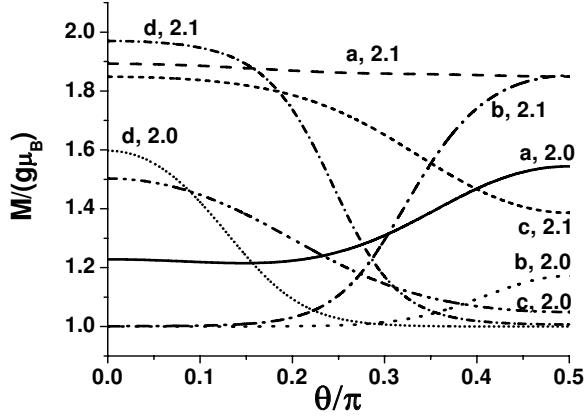}
\caption{ Plot  of $M/\gamma$ at $\phi=0$ versus $\theta/\pi$ near
the second step at $\gamma B/|J|=2.0, 2.1$ and $k_BT/|J|=0.03$ for
each $J_j/J=0.1$ with $i=a,b,c,d$. Curves are labelled with $j,
\gamma B/|J|$ values.}\label{fig24}
\end{figure}

Although not pictured explicitly, the behavior for $J_e=0.1J$ with
the other $J_j=0$ is rather  like that of the $J_a=0.1J$ curves
pictured in Figs. 20 and 21 with $\hat{\bm
z}\leftrightarrow\hat{\bm x}$, differing in ways similar to those
differences between the $J_a=0.1J$ and $J_e=0.1J$ curves pictured
for $s_1=1$ in Figs. 10-13.  As indicated in Eqs.
(\ref{B52step1})-(\ref{B52step3}), the  $B_{s,5/2}^{\rm lc}(0,0)$
are nearly independent of $J_e$. However, the $B_{s,5/2}^{\rm
lc}(\pi/2,0)+sJ$ with $J_e=0.1J$  are nearly three times as large
as for $J_a=0.1J$ case, so that the minimum  $|B_{s,5/2}^{\rm
lc}(\pi/2,0)+sJ|$ is also a minimum for $s=2$.  More details are
given in Section VII and Appendix D.

In addition, the angular dependencies of ${\bm M}$ are different
for each of the four $J_j$ we presented for $s_1=5/2$. In Fig. 24,
we present the results for $|{\bm M}(B,\theta,\phi=0)|/\gamma$
versus $\theta/\pi$ near the second level crossing at $\gamma|{\bm
B}/J|=2.0, 2.1$ and $k_BT/|J|=0.03$ for each of the four $s_1=5/2$
AFM magnetization cases pictured in Figs. 16, 18, 20, and 22. Note
that $|{\bm M}|(B,\pi-\theta,0)|=|{\bm M}(B,\theta,0)|$. The $J_a$
and $J_b$ curves, while rather similar at $\gamma|{\bm B}/J|=2.0$,
are very different at $\gamma|{\bm B}/J|=2.1$. Hence, ${\bm
M}({\bm B})$ depends strongly upon the particular type of spin
anisotropy.

\section{Analytic results for weakly anisotropic dimers of arbitrary spin}

\subsection{Induction representation eigenstates first order in the anisotropies}

Since the diagonalization of the Hamiltonian matrix is difficult
for an arbitrary magnetic field ${\bm H}$ direction and for an
arbitrary combination of spin anisotropy interactions, and must be
done for each value of $s_1$ separately, it is useful to consider
a perturbative solution in the relative strengths $J_j/J$ of the
the anisotropy interactions. We nominally assume $|J_j/J|\ll1$ for
$j=a,b,d,e$. However, to compare with low-$T$ ${\bm M}({\bm B})$
and $C_V({\bm B})$ experiments at various ${\bm H}$ directions and
magnitudes, one cannot  take $B$ to be small. In order to
incorporate an arbitrary ${\bm B}$  accurately, we rotate the
crystal axes $(\hat{\bm x}, \hat{\bm y}, \hat{\bm z})$ to
$(\hat{\bm x}', \hat{\bm y}', \hat{\bm z}')$, so that ${\bm
B}=B\hat{\bm z}'$. The rotation matrix and a brief discussion of
its ramifications are given in Appendix B.

In these rotated coordinates, the Zeeman interaction $-\gamma B
S_{z'}$ is diagonal. We therefore denote this representation as
the induction representation.   The Hamiltonian ${\cal H}'$ in
this representation is given in Appendix B.    In the induction
representation, we choose the quantum states to be
$|\varphi_s^m\rangle$.  In the absence of the four anisotropy
interactions $J_j$, ${\cal H}'={\cal H}_0'$ is diagonal,
\begin{eqnarray} {\cal
H}_0'|\varphi_s^m\rangle&=&E^{m,(0)}_{s}|\varphi_s^m\rangle,\\
\noalign{\rm where}\nonumber\\
 E^{m,(0)}_{s}&=&-Js(s+1)/2-\gamma
Bm.\label{E0}\end{eqnarray}

The operations of the remaining terms in ${\cal H}'$ on the
eigenstates $|\varphi_s^m\rangle$ are given in Appendix C.  The
first order correction to the energy in this representation is
$E_s^{m,(1)}=\langle\varphi_s^m|{\cal H}'|\varphi_s^m\rangle$,
which is found to be
\begin{eqnarray}
E_{s,s_1}^{m,(1)}&=&-\frac{J_b}{2}[2s(s+1)-1]-\frac{J_a}{2}[s(s+1)-1]\nonumber\\
& &+\frac{\tilde{J}_{b,a}^{s,s_1}}{2}[m^2+s(s+1)-1]\nonumber\\
&
&+\frac{1}{2}[s(s+1)-3m^2]\nonumber\\
& &\times\Bigl(\tilde{J}_{b,a}^{s,s_1}\cos^2\theta
+\tilde{J}_{d,e}^{s,s_1}\sin^2\theta\cos(2\phi)\Bigr),
\label{Esm1}
\end{eqnarray}
where
\begin{eqnarray}
\tilde{J}_{b,a}^{s,s_1}&=&J_b+\alpha_{s,s_1}J_a,\label{tildeJba}\\
\tilde{J}_{d,e}^{s,s_1}&=&J_d+\alpha_{s,s_1}J_e,
\label{tildeJde}\end{eqnarray} and $\alpha_{s,s_1}$ is given by
Eq. (\ref{alphass1}).

 Since the $\theta,\phi$ dependence of $E_s^{m,(1)}$ arises from
the term proportional to $\tilde{J}_{b,a}^{s,s_1}\cos^2\theta
 +\tilde{J}_{d,e}^{s,s_1}\sin^2\theta\cos(2\phi)$, it is tempting to
think that the thermodynamics with $\tilde{J}_{d,e}^{s,s_1}=0$ and
${\bm B}||\hat{\bm z}$ are equivalent to those with
$\tilde{J}_{b,a}^{s,s_1}=0$ and ${\bm B}||\hat{\bm x}$.  However,
as shown explicitly in the following, the
$\theta,\phi$-independent parts of Eq. (\ref{Esm1}) strongly break
this apparent equivalence, causing the $B_{s,s_1}^{\rm
lc}(\theta,\phi)$  for those two cases to differ.
 This
implies that $J_b$ and $J_d$ are inequivalent, as are $J_a$ and
$J_e$, even to first order in the anisotropy strengths.

\subsection{First order thermodynamics}

To first order in the anisotropy interactions, $s$ and $m$ are
still good quantum numbers, so the partition function
\begin{eqnarray}
Z&\approx&\sum_{s=0}^{2s_1}\sum_{m=-s}^se^{-\beta
E^m_{s,s_1}},\label{Zapprox}
\end{eqnarray}
where $E^m_{s,s_1}=E^{m,(0)}_{s}+E^{m,(1)}_{s,s_1}$. Although it
is difficult to perform the summation over the $m$ values
analytically,  it is nevertheless elementary to evaluate $Z$
numerically for an arbitrary $B, \theta,\phi$, and $T$ from the
eigenstate energies.     The magnetization is obtained from
\begin{eqnarray}
M(B,\theta,\phi)&\approx&\frac{\gamma}{Z}\sum_{s=0}^{2s_1}\sum_{m=-s}^sme^{-\beta
E^m_{s,s_1}}.
\end{eqnarray}

Similarly, the specific heat to first order in the anisotropy
interactions is found from
\begin{eqnarray}
C_{V}(B,\theta,\phi)&\approx&\frac{k_B\beta^2}{Z^2}\Bigl[Z\sum_{s=0}^{2s_1}\sum_{m=-s}^s(E^m_{s,s_1})^2e^{-\beta
E^m_{s,s_1}}\nonumber\\
& &-\Bigl(\sum_{s=0}^{2s_1}\sum_{m=-s}^sE^m_{s,s_1}e^{-\beta
E^m_{s,s_1}}\Bigr)^2\Bigr].\label{CVapprox}
\end{eqnarray}

\begin{figure}
\includegraphics[width=0.45\textwidth]{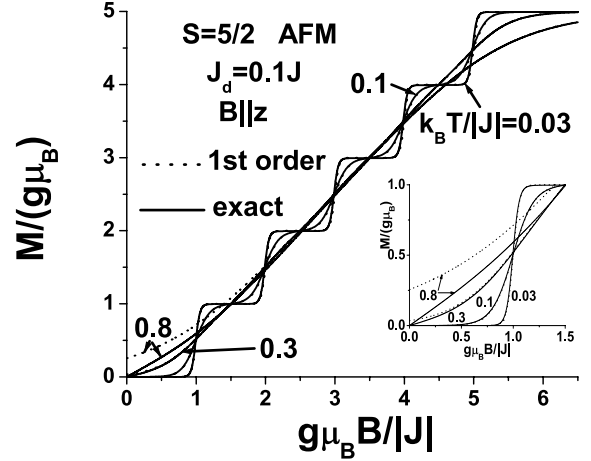}
\caption{Comparison of $M/\gamma$ versus $\gamma B/|J|$ obtained
using the first-order asymptotic form (dotted) with the exact
calculation (solid), for the $s_1=5/2$ AFM dimer with $J_d=0.1J$,
$J_a=J_b=J_e=0$, at $k_BT/|J|=0.03,0.1,0.3,0.8$, as indicated.
Inset:  expanded view of the region $0\le \gamma B/|J|\le1.5$.}
\label{magtest}
\end{figure}

\begin{figure}
\includegraphics[width=0.45\textwidth]{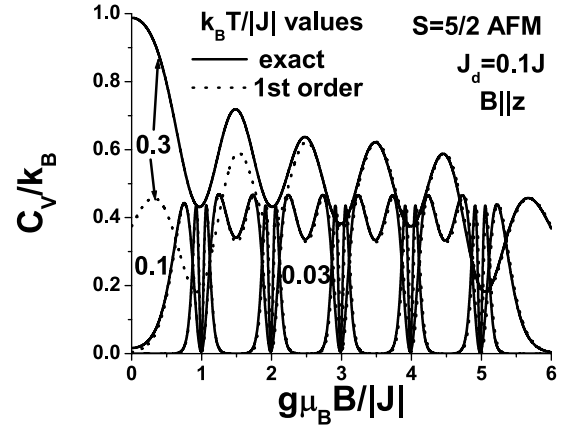}
\caption{Comparison of $C_V/k_B$ versus $\gamma B/|J|$ obtained
using the first-order asymptotic form (dotted) with the exact
calculation (solid), for the $s_1=5/2$ AFM dimer with $J_d=0.1J$,
$J_a=J_b=J_e=0$, at $k_BT/|J|=0.03,0.1,0.3$, as indicated.}
\label{shtest}
\end{figure}

As a test of the accuracy of this first-order calculation, we have
compared the first-order and exact $M(B)$  obtained for the
$s_1=5/2$ dimer with $J_d=0.1J$ and ${\bm B}||\hat{\bm z}$ in Fig.
\ref{magtest}.  The corresponding comparison between the
first-order and exact $C_V(B)$  is shown in Fig. \ref{shtest}. We
see that the  curves evaluated using the first-order and the exact
expressions for $M$ and $C_V$ with $s_1=5/2$ are indistinguishable
at $k_BT/|J|=0.03$. The $C_V$ curves are noticeably different at
$k_BT/|J|=0.1$ for $\gamma B/|J|<0.4$, and at $k_BT/|J|=0.3$, they
are noticeably different for $\gamma B/|J|<2.6$. Corresponding
noticeable differences in the $M$ curves at the same $B$ values
appear at $T$ values roughly three times as high as in the $C_V$
curves.

At very low $T$, $k_BT/|J|\ll1$, the most important states in this
perturbative scheme are the minima for each $s$ value,
$E_{s,s_1}^s$, which determine the first-order level crossings. As
$T\rightarrow0$, we can ignore all of the  $m\ne s$ states in Eqs.
(\ref{Zapprox})-(\ref{CVapprox}).

\subsection{Level crossings first order in the anisotropies}

We can find an expression for the $s^{\rm th}$ AFM level crossing
at the induction $B^{{\rm lc}(1)}_{s,s_1}$ to first order in the
anisotropy interactions for a general $s_1$ spin dimer by equating
$E^{s,(0)}_{s}+E^{s,(1)}_{s,s_1}$ to
$E^{s-1,(0)}_{s-1}+E^{s-1,(1)}_{s-1,s_1}$,
\begin{eqnarray}\gamma B^{{\rm lc}(1)}_{s,s_1}&=&-Js-J_b/2-c_{s,s_1}J_a\nonumber\\
& &-\frac{(4s-3)}{2}[J_b\cos^2\theta+J_d\sin^2\theta\cos(2\phi)]\nonumber\\
& &+3c_{s,s_1}[J_a\cos^2\theta+J_e\sin^2\theta\cos(2\phi)],\\
c_{s,s_1}&=&\frac{[3+3s-5s^2-4s^3+4s_1(s_1+1)]}{2(2s+1)(2s+3)}.\label{Bstep}
\end{eqnarray}
This expression is consistent with those obtained for $s_1=1/2,1$
given by Eqs. (\ref{Bstephalf}) and (\ref{b11z})-(\ref{b21xy}). In
addition, this expression is nearly quantitatively in agreement
with the $M(B)$ and $C_V(B)$ behaviors pictured for $s_1=5/2$ in
Figs. 16-23. We note that $\gamma B_{s,s_1}^{{\rm lc}(1)}$
contains the $\theta,\phi$-independent terms,
$-Js-J_b/2-c_{s,s_1}J_a$, which distinguish $J_b$ from $J_d$ and
$J_a$ from $J_e$.

 In particular, we note
that the single-ion anisotropy interactions behave very
differently with increasing step number than do the global
anisotropy interactions, especially for large $s_1$. For the three
cases we studied in detail, for $s_1=1/2$, $c_{1,1/2}=0$, so that
the local anisotropy terms are irrelevant, for $s_1=1$,
$c_{1,1}=\frac{1}{6}$ and  $c_{2,1}=-\frac{1}{2}$ have different
signs, and for $s_1=5/2$ as in Fe$_2$ dimers, the first three
$c_{s,5/2}$ coefficients are $\frac{16}{15}$, $-\frac{4}{35}$, and
$-\frac{53}{63}$, respectively, the second being an order of
magnitude smaller than the other two, and opposite in sign from
the first. For $s_1=9/2$ dimers such as [Mn$_4$]$_2$, the first
four $c_{s,9/2}$ are $\frac{16}{5}$, $\frac{4}{5}$,
$-\frac{1}{3}$, and $-\frac{37}{33}$, which changes sign between
$s=2$ and $s=3$, where its magnitude is a minimum.

This is in sharp contrast to the global anisotropy interactions,
for which the analogous coefficient $(4s-3)/2$ increases
monotonically from $\frac{1}{2}$ to $\frac{5}{2}$ as $s$ increases
from 1 to 2, independent of $s_1$. These differences should be
possible to verify experimentally in careful low-$T$ experiments
at high magnetic fields applied at various directions on single
crystals of those $s_1=5/2$ Fe$_2$ and $s_1=9/2$ [Mn$_4$]$_2$
dimers for which $|J|$ is sufficiently small.

\subsection{Level crossings to second order in the anisotropy energies}

 To aid in the
analysis of experimental data, we have extended this perturbative
calculation to second order.  Since we expect the $|J_j/J|\le0.1$
in most circumstances, this extension ought to be sufficient to
accurately analyze most experimentally important samples. To
second order in the anisotropy interactions, the eigenstate
energies $E_{s,s_1}^{m,(2)}$ are given in Appendix C. We note that
the $E_{s,s_1}^{m(2)}$ contains divergences at $\gamma B/|J|=0,
2s-1, 2s+3, s-1/2,$ and $s+3/2$, so that near to those values, one
would need to modify the perturbation expansion to  take proper
account of the degeneracies. Hence, the expressions for
$E_{s,s_1}^{m(2)}$ cannot be used in the asymptotic expressions
for the thermodynamics, Eqs. (\ref{Zapprox})-(\ref{CVapprox}).
However, as the $s$th AFM level crossing occurs approximately at
$\gamma B/|J|=s$, which is far from any divergences, we can safely
use this second order expansion to obtain an expression for the
level crossings second order in the anisotropy interaction
energies.  We find
\begin{eqnarray}
\gamma B^{{\rm lc}(2)}_{s,s_1}&=&\gamma B^{{\rm
lc}(1)}_{s,s_1}+\Bigl(E^{s,(2)}_{s,s_1}-E_{s-1,s_1}^{s-1,(2)}\Bigr)\Bigr|_{B=-Js/\gamma},\label{Bstep2}\nonumber\\
\end{eqnarray}
where $\gamma B^{{\rm lc}(1)}_{s,s_1}$ is given by Eq.
(\ref{Bstep}).

The full expression for $B_{s,s_1}^{{\rm lc}(2)}$ is given in
Appendix D.  From this expression, it is easy to see that for
$s_1=1/2$, $E_{1,1/2}^{{\rm
lc}(2)}=\frac{1}{2}f_1(\theta,\phi)+\frac{1}{8}f_4(\theta,\phi)$,
where $f_1$ and $f_4$ are given in Appendix D.  For ${\bm
B}||\hat{\bm z}$, $E_{1,1/2}^{{\rm lc}(2)}=-\frac{1}{2}J_d^2/|J|$,
and with ${\bm B}||\hat{\bm x},\hat{\bm y}$, $E_{1,1/2}^{{\rm
lc}(2)}=-\frac{1}{8}(J_b\pm J_d)^2/|J|$, in agreement with the
expansion to second order in the $J_j$ of our exact formulas in
Eq. (\ref{Bstephalf}).  However, the second order functions have
more complicated $\theta,\phi$ dependencies than do the first
order $B^{{\rm lc}(1)}_{s,s_1}$ in Eq. (\ref{Bstep}).  We have
also explicitly checked each formula in Appendix D for $s_1=1$ and
$s=1,2$.  Thus, Eq. (\ref{Bstep2}) is a highly accurate expression
for the full $\theta,\phi$ dependencies of all $s=1,\ldots,2s_1+1$
level crossings of a single crystal dimer of single ion spin
$s_1$.

 By superposing this non-universal level-crossing  formula, Eq.
 (\ref{Bstep2}),
combined with the universal behavior presented in Eqs.
(\ref{CVzereos})-(\ref{slope}), it is easy to use our results in
accurate fits to experimental data at low temperatures and high
magnetic fields on dimers with arbitrary $s_1$ values.

\section{Summary and conclusions}

In summary, we solved for the low-temperature  magnetization and
specific heat of equal spin $s_1$ antiferromagnetic dimer single
molecule magnets, including the most general forms of anisotropic
spin exchange interactions quadratic in the spin operators. The
magnetization and specific heat  exhibit steps and zeroes,
respectively, at the non-universal level-crossing induction values
$B_{s,s_1}^{\rm lc}(\theta,\phi)$, but the magnetization steps and
their midpoint slopes, plus the two peaks surrounding the specific
heat zeroes all exhibit universal behavior at sufficiently low
temperatures. Local (or single-ion) anisotropy interactions lead
to low-temperature magnetization step plateaus that have a much
richer variation with the magnetic induction ${\bm B}$ than do
those obtained from global anisotropy interactions, provided that
 $s_1>1/2$. We derived simple, accurate
asymptotic analytic expressions for the low-temperature
magnetization and specific heat for the most general quadratic
anisotropic spin interactions at an arbitrary ${\bm B}$, and an
accurate expression for $B_{s,s_1}^{\rm lc}(\theta,\phi)$,
enabling fast and accurate fits to experimental data.

There were two low-$T$ $M(B)$ studies of Fe$_2$
dimers.\cite{Fe2mag,Fe2Cl}  For
$\mu$-oxalatotetrakis(acetylacetonato)Fe$_2$, all five peaks in
$dM/dH$ were measured in pulsed magnetic fields $H$.  These evenly
spaced peaks
  indicated little, if any, spin
anisotropy effects.\cite{Fe2mag}  On the other hand, studies of
the first 2-3 $dM/dH$ peaks in powdered samples of
[Fe(salen)Cl]$_2$, where salen is
$N,N'$-ethylenebis(salicylideneiminato), were much more
interesting.\cite{Fe2Cl} These data showed a broad first peak at
$B=17-20$T that was only partially resolvable into two separate
peaks, followed by a sharp second peak at $B=36$ T, consistent
with  local axial anisotropy of strength $|J_a/J|\approx0.1$, as
 obtained from the derivatives of the curves shown in Fig. 20
 From Eqs.
 (\ref{B52step1})-(\ref{B52step3}), one could also have $|(J_a\pm
3J_3)/J|\approx0.1$.
 These values might perhaps be combined with a smaller
$|J_c/J|$, obtained from the derivatives of the curves
 pictured in Fig. 22.

 Without a detailed single
 crystal study with the magnetic field directed along a number of different crystal
 directions, it is impossible to determine the relative amounts of
 $J_a$ and $J_e$ that would best fit the data.
However, the existing data on [Fe(salen)Cl]$_2$ appear to be
inconsistent with a predominant global anisotropy interaction of
either type, as obtained from the derivatives of the curves shown
in Figs. 16 and 18. Only single crystal studies could determine if
a small amount of such global anisotropy interactions were present
in addition to one or more presumably larger local spin anisotropy
interactions. In comparing the two materials cited above, it
appears that the interaction of a Cl$^{-}$ ion neighboring each
Fe$^{3+}$ ion leads to strong local (or single-ion) anisotropy
effects.  In order to verify this hypothesis and to elucidate the
details of the interactions, further experiments using single
crystals in different field orientations on this and related
Fe$_2$ dimers with 1-3 similarly bonded Cl$^{-}$ ions are
urged.\cite{Fe2Cl3,Fe2Clnew}  We also urge single crystal data on
some of the $s_1-9/2$ [Mn$_4$]$_2$
dimers,\cite{Mn4dimer,Mn4dimerDalal} as well as on $s_1=1/2$
dimers lacking in predicted local spin anisotropy effects.
\cite{V2neutron,Gudel,V2P2O9,ek} To aid in the fits, we derived
simple, useful formulas for the magnetization and specific heat at
low temperature and sufficiently large magnetic induction, and
accurate formulas for the level crossing inductions.

With local anisotropy interactions, the total spin $s$ is not a
good quantum number, potentially modifying our understanding of
quantum tunneling processes. It might also be possible to fit a
variety of experimental results using a smaller, consistent set of
model parameters.\cite{Fe8spin9} We emphasize that the study of
SMM dimers, for which the most general anisotropic quadratic
exchange interactions can be solved exactly, may be our best hope
for attaining a more fundamental understanding of the underlying
physics of single molecule magnets.

\section{Acknowledgments}

We thank the Max-Planck-Institut f{\"u}r Physik komplexer Systeme,
Dresden, Germany, the University of North Dakota, Grand Forks, ND,
USA, and Talat S. Rahman for their kind hospitality and support.
This work was supported by the Netherlands Foundation for the
Fundamental Research of Matter and by the NSF under contract
NER-0304665.

\section{Appendix A}
\subsection{Specific heat details for $s_1=1/2$}
We first present the numerators of the exact expressions for the
specific heat with $s_1=1/2$  and ${\bm B}||\hat{\bm i}$ for
$i=x,y,z$.  We have
\begin{eqnarray}
{\cal N}_{x,y}&=&F_{x,y}^2+\frac{1}{4}(J+2J_{y,x})^2e^{\beta[2(J_{y,x}-J_{x,y})-J]}\nonumber\\
& &+F_{x,y}\sinh(\beta F_{x,y})e^{-\beta
J_{x,y}}\nonumber\\
& &\times[(J+J_{x,y})e^{-\beta J}+(J_{x,y}-2J_{y,x})e^{2\beta
J_{y,x}}]\nonumber\\
& &+\frac{1}{2}e^{-\beta J_{x,y}}\cosh(\beta
F_{x,y})\nonumber\\
& &\times\Bigl([(J+J_{x,y})^2+F_{x,y}^2]e^{-\beta
J}\nonumber\\
& &+[(2J_{y,x}-J_{x,y})^2+F_{x,y}^2]e^{2\beta J_{y,x}}\Bigr),\\
 {\cal N}_z&=&F_z^2\cosh(2\beta
F_z)+\frac{J^2}{4}e^{-\beta(J+2J_b)}\nonumber\\
& &+\frac{1}{2}\cosh(\beta
F_z)\Bigl(\Delta_z(J_b^2+F_z^2)\nonumber\\
& &\qquad+e^{-\beta(J+J_b)}J(J+2J_b)\Bigr)\nonumber\\
& &+F_z\sinh(\beta F_z)\Bigl(J_b\Delta_z+Je^{-\beta(J+J_b)}\Bigr).
\end{eqnarray}

\subsection{Eigenvalues for $s_1=1$}

In the remainder of this appendix, we provide the details of our
exact results for $s_1=1$. The cubic equation for the three $s=1$
eigenvalues is given by
\begin{eqnarray}
\epsilon_n&=&-J-J_b-J_a+\lambda_n,\>\>{\rm for}\>\>n=2,3,4,\\\
 0&=&
-\lambda_n^3-(J_a-J_b)\lambda_n^2+\lambda_n[b^2+(J_d-J_e)^2]\nonumber\\
& &+(J_a-J_b)[b_z^2+(J_d-J_e)^2]\nonumber\\
& &-(J_d-J_e)(b_x^2-b_y^2),
\end{eqnarray}
where the $b_i=\gamma B_i$ as in Eq. (\ref{bi}).
 For  ${\bm B}||\hat{\bm
z}$, the cubic equation is easily solved to yield
\begin{eqnarray}
\lambda_n&=&J_b-J_a,\pm\sqrt{b^2+(J_d-J_e)^2}.
\end{eqnarray}
For  ${\bm B}||\hat{\bm x},\hat{\bm y}$, we have
\begin{eqnarray}
\lambda_n&=&\pm(J_e-J_d),-J_{y,x}\pm\sqrt{b^2+\overline{J}_{x,y}^2},\\
\overline{J}_{x,y}&=&\frac{1}{2}[J_a-J_b\mp(J_d-J_e)],
\end{eqnarray}
where $\overline{J}_x$ ($\overline{J}_y$) corresponds to the upper
(lower) sign.
 The  rank ordering of these eigenvalues depends
upon $B$ and the various anisotropy parameters, of course.

 The six eigenvalues for
the mixed $s=0,2$ states satisfy
\begin{eqnarray}
\epsilon_n&=&-3J-\frac{4}{3}J_a-2J_b+\lambda_n\label{epsilonnspin1}
\end{eqnarray}
 We then define
\begin{eqnarray}
\tilde{J}_a&=&\frac{\sqrt{8}}{3}J_a,\label{Jatilde}\\
\tilde{J}_b&=&J_b+J_a/3,\label{Jbtilde}\\
\tilde{J}_d&=&J_d+J_e/3,\label{Jdtilde}\\
\tilde{J}_e&=&\frac{2}{\sqrt{3}}J_e,\label{Jetilde}\\
\tilde{J}&=&J-\frac{2}{9}J_a,\label{Jtilde}
\end{eqnarray}
and obtain the Hermitian matrix $\tensor{\bm M}$ for the six
$s=0,2$ states, the eigenvalues of which are the $\lambda_n$. For
brevity, we let
\begin{eqnarray}
Q_n^p&=&n\tilde{J}_b+pb\cos\theta,\\
b_{\perp}&=&b\sin\theta e^{-i\phi},\\
b_3&=&\sqrt{\frac{3}{2}}b_{\perp},\\
a&=&\tilde{J}_a,\\
d_3&=&3\tilde{J}_d,\\
d_6&=&\sqrt{6}\tilde{J}_d,\\
e&=&\tilde{J}_e.
\end{eqnarray}
Then, the matrix $\tensor{\bm M}$ is given by
\begin{eqnarray}
\tensor{\bm M}\!\!&=&\!\!\left(\begin{array}{cccccc}Q_{-2}^{-2}&-b_{\perp}&-d_6&0&0&-e\\
-b_{\perp}^{*}&Q_1^{-1}&-b_3&-d_3&0&0\\
-d_6&-b_3^{*}&Q_2^{0}&-b_3&-d_6&-a\\
0&-d_3&-b_3^{*}&Q_1^1&-b_{\perp}&0\\
0&0&-d_6&-b_{\perp}^{*}&Q_{-2}^2&-e\\
-e&0&-a&0&-e&Q_2^0+3\tilde{J}\end{array}\right).
\end{eqnarray}

After some rearrangement, the $\lambda_n$ are found to satisfy
\begin{eqnarray}
0&=&\sum_{p=0}^6c_p(\lambda_n)^p,\label{hexatic}\\
c_6&=&1,\label{c6}\\
c_5&=&-2\tilde{J}_b-3\tilde{J},\\
c_4&=&-\tilde{J}_a^2-7\tilde{J}_b^2-21\tilde{J}_d^2-2\tilde{J}_e^2-5b^2,\label{c4}\\
c_3&=&3\tilde{J}(7\tilde{J}_b^2+21\tilde{J}_d^2+5b^2)\nonumber\\
& &+2\tilde{J}_b(-\tilde{J}_a^2+8\tilde{J}_b^2+12\tilde{J}_d^2+2\tilde{J}_e^2)\nonumber\\
& &+4\sqrt{6}\tilde{J}_a\tilde{J}_d\tilde{J}_e+21\tilde{J}_d(b_x^2-b_y^2)\nonumber\\
& &+3\tilde{J}_b(b^2+7b_z^2),\\
c_2&=&3\tilde{J}\Bigl(\tilde{J}_b[-2\tilde{J}_b^2+18\tilde{J}_d^2+7(b^2-3b_z^2)]\nonumber\\
& &\qquad-21\tilde{J}_d(b_x^2-b_y^2)\Bigr)\nonumber\\
&
&+\tilde{J}_a^2(3\tilde{J}_b^2+9\tilde{J}_d^2+2b^2+3b_z^2)\nonumber\\
&
&+2\tilde{J}_b^2[4\tilde{J}_b^2+3\tilde{J}_e^2+11(b^2-3b_z^2)]\nonumber\\
&
&+2\tilde{J}_d^2[54(\tilde{J}_b^2+\tilde{J}_d^2)+9\tilde{J}_e^2-12(b^2-3b_z^2)]\nonumber\\
& &+2\tilde{J}_e^2(4b^2-3b_z^2)+4b^4\nonumber\\
&
&+2(3\tilde{J}_b\tilde{J}_d-\sqrt{6}\tilde{J}_a\tilde{J}_e)(b_x^2-b_y^2),\\
c_1&=&-12\tilde{J}\Bigl(b^4+3\tilde{J}_b^4+18\tilde{J}_b^2\tilde{J}_d^2+27\tilde{J}_d^4\nonumber\\
& &+2(\tilde{J}_b^2-3\tilde{J}_d^2)(b^2-3b_z^2)\nonumber\\
&
&+12\tilde{J}_b\tilde{J}_d(b_x^2-b_y^2)\Bigr)\nonumber\\
&
&-2\tilde{J}_b\Bigl(2(9\tilde{J}_d^2+b^2)(b^2+3b_z^2)\nonumber\\
&
&-\tilde{J}_a^2(b^2-3b_z^2)+6\tilde{J}_b^2(b^2-5b_z^2)\nonumber\\
&
&+(42\tilde{J}_b\tilde{J}_d+\sqrt{6}\tilde{J}_a\tilde{J}_e)(b_x^2-b_y^2)\nonumber\\
&
&-2\tilde{J}_a^2(\tilde{J}_b^2+9\tilde{J}_d^2)+8\tilde{J}_b^2(2\tilde{J}_b^2+6\tilde{J}_d^2+\tilde{J}_e^2)\nonumber\\
&
&+6\sqrt{6}\tilde{J}_a\tilde{J}_b\tilde{J}_d\tilde{J}_e\Bigr)\nonumber\\
&
&-2\tilde{J}_d\Bigl(6(b^2-3\tilde{J}_d^2+\tilde{J}_e^2)(b_x^2-b_y^2)\nonumber\\
&
&-\sqrt{6}\tilde{J}_a\tilde{J}_e(b^2-3b_z^2-18\tilde{J}_d^2)\Bigr),\\
c_0&=&-12\tilde{J}\Bigl(\tilde{J}_bb^2(b^2-3b_z^2)\nonumber\\
& &-3\tilde{J}_d(b^2-\tilde{J}_b^2-3\tilde{J}_d^2)(b_x^2-b_y^2)\nonumber\\
&
&+\tilde{J}_b^3(b^2+3b_z^2-2\tilde{J}_b^2+12\tilde{J}_d^2)\nonumber\\
&
&-3\tilde{J}_b\tilde{J}_d^2(7b^2-3b_z^2-18\tilde{J}_d^2)\Bigr)\nonumber\\
&
&-\tilde{J}_a^2\Bigl((b^2-3b_z^2+2\tilde{J}_b^2)^2+36\tilde{J}_d^2(b_z^2-\tilde{J}_b^2)\Bigr)\nonumber\\
&
&-4\tilde{J}_b^2\Bigl(2b^2(b^2-3b_z^2+\tilde{J}_e^2)-2\tilde{J}_b^2(2\tilde{J}_b^2+\tilde{J}_e^2-b^2-3b_z^2)\nonumber\\
&
&+6\tilde{J}_d^2(-7b^2+3b_z^2+4\tilde{J}_b^2+18\tilde{J}_d^2+3\tilde{J}_e^2)\nonumber\\
&
&+(6\tilde{J}_b\tilde{J}_d-\sqrt{6}\tilde{J}_a\tilde{J}_e)(b_x^2-b_y^2)\Bigr)\nonumber\\
&
&+24\tilde{J}_b\tilde{J}_d(b_x^2-b_y^2)(b^2-3\tilde{J}_d^2-2\tilde{J}_e^2)\nonumber\\
&
&+2\sqrt{6}\tilde{J}_a\tilde{J}_e(b_x^2-b_y^2)(b^2-3b_z^2-6\tilde{J}_d^2)\nonumber\\
& &-6\tilde{J}_e^2(b_x^2-b_y^2)^2\nonumber\\
&
&+8\sqrt{6}\tilde{J}_a\tilde{J}_b\tilde{J}_d\tilde{J}_e(2b^2-3b_z^2+\tilde{J}_b^2-9\tilde{J}_d^2).
\end{eqnarray}

\subsection{${\bm B}$ along a crystal axis}

We recall that the six Hamiltonian matrix eigenvalues $\epsilon_n$
are generally given by Eq. (\ref{epsilonnspin1}).  For the special
case ${\bm B}||\hat{\bm z}$, $\tensor{\bm M}$ is block diagonal.
The resulting eigenvalues of the nominal $s=2,m=\pm1$ notation are
\begin{eqnarray}
\lambda_{6,8}^z&=&\tilde{J}_b\mp\sqrt{b^2+9\tilde{J}_d^2}.
\end{eqnarray}
The remaining four states are obtained from
\begin{eqnarray}
0&=&\sum_{p=0}^4k_p^z(\lambda_n^z)^p,\\
k_4^z&=&1,\\
k_3^z&=&-3\tilde{J},\\
k_2^z&=&-6\tilde{J}\tilde{J}_b-\tilde{J}_a^2-8\tilde{J}_b^2-12\tilde{J}_d^2-2\tilde{J}_e^2-4b^2,\\
k_1^z&=&12\tilde{J}(b^2+\tilde{J}_b^2+3\tilde{J}_d^2)\nonumber\\
&
&+4\tilde{J}_b(4b^2-\tilde{J}_a^2)+4\sqrt{6}\tilde{J}_a\tilde{J}_d\tilde{J}_e,\\
k_0^z&=&24\tilde{J}\tilde{J}_b(-b^2+\tilde{J}_b^2+3\tilde{J}_d^2)+4\tilde{J}_a^2(b^2-\tilde{J}_b^2)\nonumber\\
&
&+8\tilde{J}_b^2(-2b^2+2\tilde{J}_b^2+6\tilde{J}_d^2+\tilde{J}_e^2)\nonumber\\
& &+8\sqrt{6}\tilde{J}_a\tilde{J}_b\tilde{J}_d\tilde{J}_e.
\end{eqnarray}

Similarly, for ${\bm B}||\hat{\bm x},\hat{\bm y}$, $\tensor{\bm
M}$ is also block diagonal.  The resulting eigenvalues of the
nominal $s=2, m=\pm1$ notation are
\begin{eqnarray}
\lambda_{6,8}^{x,y}&=&\frac{3}{2}\tilde{J}_{x,y}-2\tilde{J}_b\mp\sqrt{b^2+\tilde{J}_{x,y}^2},\\
\tilde{J}_{x,y}&=&\frac{3}{2}(\tilde{J}_b\pm\tilde{J}_d).\label{tildeJxy}
\end{eqnarray}
The remaining four states are obtained from
\begin{eqnarray}
0&=&\sum_{p=0}^4n^{x,y}_p(\lambda_n^{x,y})^p,\\
n^{x,y}_4&=&1,\\
n^{x,y}_3&=&-3(\tilde{J}+\tilde{J}_b\mp\tilde{J}_d),\\
n^{x,y}_2&=&-\tilde{J}_a^2-2\tilde{J}_b^2-12\tilde{J}_d^2-2\tilde{J}_e^2+3\tilde{J}\tilde{J}_b\nonumber\\
& &-4b^2\mp3\tilde{J}_d(2\tilde{J}_b+3\tilde{J}),\\
n^{x,y}_1&=&6\tilde{J}_e^2(\tilde{J}_b\mp\tilde{J}_d)+4b^2(3\tilde{J}+\tilde{J}_b\pm3\tilde{J}_d)\nonumber\\
&
&+12(\tilde{J}_b^2+3\tilde{J}_d^2)(\tilde{J}+\tilde{J}_b\mp\tilde{J}_d)\nonumber\\
& &-\tilde{J}_a^2(\tilde{J}_b\pm3\tilde{J}_d)+4\sqrt{6}\tilde{J}_a\tilde{J}_d\tilde{J}_e,\\
n^{x,y}_0&=&b^2[\tilde{J}_a^2+12\tilde{J}(\tilde{J}_b\mp3\tilde{J}_d)+8\tilde{J}_b^2+6\tilde{J}_e^2\nonumber\\
&
&\mp24\tilde{J}_b\tilde{J}_d\mp2\sqrt{6}\tilde{J}_a\tilde{J}_e] \nonumber\\
&
&+2\tilde{J}_b^2[\tilde{J}_a^2-4\tilde{J}_b^2-6\tilde{J}\tilde{J}_b-12\tilde{J}_d^2-2\tilde{J}_e^2]\nonumber\\
& &\pm12\tilde{J}_d(2\tilde{J}_b+3\tilde{J})(\tilde{J}_b^2+3\tilde{J}_d^2)\nonumber\\
&
&\mp6\tilde{J}_b\tilde{J}_d(\tilde{J}_a^2\pm6\tilde{J}\tilde{J}_d-2\tilde{J}_e^2)\nonumber\\
&
&-4\sqrt{6}\tilde{J}_a\tilde{J}_d\tilde{J}_e(\tilde{J}_b\mp3\tilde{J}_d).
\end{eqnarray}

\subsection{Simple special cases}

When only one of the $J_j\ne0$, the eigenvalues for ${\bm
B}||\hat{\bm z}$ simplify considerably.  For $J_b\ne0$, we have
\begin{eqnarray}
\lambda_n^z&=&3J+2J_b, 2J_b, J_b\pm b, -2J_b\pm 2b.
\end{eqnarray}
For $J_d\ne0$, we find
\begin{eqnarray}
\lambda_n^z&=&0,3J, \pm\sqrt{b^2+9J_d^2}, \pm2\sqrt{b^2+3J_d^2}.
\end{eqnarray}
For $J_a\ne0$, the eigenvalues can also be found analytically,
\begin{eqnarray}
\lambda_n^z&=&-\frac{2J_a}{3}\pm 2b, \frac{J_a}{3}\pm
b,\frac{J_a}{3}+\frac{3J}{2}\pm\sqrt{\frac{9}{4}J^2+J_a^2-JJ_a}.\nonumber\\
\end{eqnarray}
We note that the ground state energy in this case is
\begin{eqnarray}
E_1&=&-\frac{3}{2}J-J_a-\sqrt{\frac{9}{4}J^2+J_a^2-JJ_a},
\end{eqnarray}
which explicitly involves mixing of the $s=m=0$ and the $s=2, m=0$
states.  For $J_e\ne0$, three of the eigenvalues are
\begin{eqnarray}
\lambda_n^z&=&0,\pm\sqrt{b^2+J_e^2},\\
\end{eqnarray}
and the remaining three satisfy the cubic equation,
\begin{eqnarray}
0&=&x^3-3Jx^2-4x(b^2+J_e^2)+4J(J_e^2+3b^2).
\end{eqnarray}
We note that the cubic equation must be solved to obtain the
ground state energy, since the $s=m=0$ state mixes with the
$s=2,\pm2$ states.  For ${\bm B}||\hat{\bm x},\hat{\bm y}$, with
$J_b\ne0, J_d\ne0$, the eigenvalues are
\begin{eqnarray}
\lambda_n^{x,y}&=&3J+2J_b,-2J_b+\tilde{J}_{x,y}\pm\sqrt{b^2+\tilde{J}_{x,y}^2},
\end{eqnarray}
where $\tilde{J}_{x,y}$ is given by Eq. (\ref{tildeJxy}), and the
other three eigenvalues satisfy the cubic equation,
\begin{eqnarray}
0&=&-x^3+(J_b\mp3J_d)x^2+4(b^2+J_b^2+3J_d^2)x\nonumber\\
& &+4(b^2-J_b^2-3J_d^2)(J_b\mp3J_d).
\end{eqnarray}
For $J_a\ne0$, two of the eigenvalues satisfy
\begin{eqnarray}
\lambda_n^{x,y}&=&-\frac{J_a}{6}\pm\sqrt{b^2+J_a^2/4},
\end{eqnarray}
 and the other four eigenvalues satisfy the quartic equation,
 \begin{eqnarray}
 0&=&81x^4-27(J_a+9J)x^3-9(36b^2-9JJ_a+12J_a^2)x^2\nonumber\\
 & &+12(-J_a^3+9JJ_a^2-9b^2J_a+81b^2J)x\nonumber\\
 & &+4(4J_a^4-9JJ_a^3+18b^2J_a^2+81b^2JJ_a).
 \end{eqnarray}
 Finally, for $J_e\ne0$, two eigenvalues are
 \begin{eqnarray}
 \lambda_n^{x,y}&=&\pm\frac{J_e}{2}+\sqrt{b^2+J_e^2/4},
 \pm\frac{J_e}{2}-\sqrt{b^2+J_e^2/4},\nonumber\\
 \end{eqnarray}
 and the other four eigenvalues satisfy the quartic equation,
 \begin{eqnarray}
 0&=&x^4-(3J\mp J_e)x^3-(4b^2+4J_e^2\pm3JJ_e)x^2\nonumber\\
 & &+4[3b^2J+JJ_e^2\pm J_e(b^2-J_e^2)]x\nonumber\\
 & &\pm4J_e(-3b^2J\pm2b^2J_e+JJ_e^2).
 \end{eqnarray}

  At the first level crossing with ${\bm
B}||\hat{\bm z}$, we  have for $J_a\ne0$,
\begin{eqnarray} \gamma B_{1,1}^{{\rm
lc},z}&=&\frac{1}{2}\Bigl(J+[9J^2+4J_a^2-4JJ_a]^{1/2}\Bigr).\label{Jastep1}
\end{eqnarray}
At the second level crossing, simple formulas are only obtained
for ${\bm B}||\hat{\bm z}$ with  one $J_j\ne0$. For ${\bm
B}||\hat{\bm z}$ and $J_a\ne0$, $J_b\ne0$ and $J_d\ne0$,
respectively, we have
\begin{eqnarray}
\gamma B_{2,1}^{{\rm
lc},z}&=&\left\{\begin{array}{c}-2J-J_a,\\
-2J -3J_b,\\
\\
\frac{1}{3}\Bigl(20J^2-30J_d^2+8[4J^4-6J^2J_d^2]^{1/2}\Bigr)^{1/2}.\end{array}\right.\label{secondstep}
\end{eqnarray}

\section{Appendix B}
\subsection{Rotation to the induction representation}
The rotation from the crystal representation to the induction
representation is obtained from
\begin{eqnarray}
\left(\begin{array}{c}\hat{\bm x}\\
\hat{\bm y}\\
\hat{\bm z}\end{array}\right)&=&\left(\begin{array}{ccc}\cos\theta\cos\phi &-\sin\phi &\sin\theta\cos\phi\\
\cos\theta\sin\phi &\cos\phi &\sin\theta\sin\phi\\
-\sin\theta
&0&\cos\theta\end{array}\right)\left(\begin{array}{c}\hat{\bm
x}'\\
\hat{\bm y}'\\
\hat{\bm z}'\end{array}\right),\nonumber\\
\end{eqnarray}
leading to ${\bm B}=B\hat{\bm z}'$.\cite{klemm}  This operation is
equivalent to a rotation by $-\pi/2$ about the $z$ axis, a
rotation by $\theta$ about the transformed $x$ axis, and then a
rotation by $\pi/2-\phi$ about the transformed $z$
axis.\cite{Goldstein}  In effect, in using the above rotation
matrix, we made the arbitrary choice that the rotated $z$ axis
lies in the $x'z'$ plane. After the above rotation, it is still
possible to rotate the crystal by an arbitrary angle $\chi$ about
the $z'$ axis, keeping ${\bm B}||\hat{\bm z}'$.  Hence, there are
in effect an infinite number of equivalent rotations leading to
${\bm B}=B\hat{\bm z}'$.  The resulting Hamiltonian matrix will
then have off-diagonal elements that depend upon $\chi$.
 However, all such rotations necessarily lead to the identical, $\chi$-independent, set
of eigenvalues of the resulting diagonalized Hamiltonian matrix.
We have explicitly checked that the above rotation gives the exact
cubic expression, Eq. (\ref{halfeigen}), for the $s_1=1/2$
eigenvalues, and also leads to the correct eigenstate energies
second order in each of the $J_j$ for $s_1=1$.   We also showed
explicitly that $\chi$ does not enter the eigenstate energies
second order in $J_b$ for arbitrary $s, s_1, m$.

\subsection{Global Hamiltonian}
The global axial and azimuthal anisotropy interactions in the
rotated frame are
\begin{eqnarray}{\cal
H}_b'&=&-J_b\Bigl(S_{z'}^2\cos^2\theta+S_{x'}^2\sin^2\theta\nonumber\\
& &\qquad-\sin(2\theta)\{S_{x'},S_{z'}\}/2\Bigr),\\
{\cal
H}_d'&=&-J_d\Bigl[\cos(2\phi)\Bigl(S_{x'}^2\cos^2\theta+S_{z'}^2\sin^2\theta-S_{y'}^2\nonumber\\
&
&\qquad+\sin(2\theta)\{S_{x'},S_{z'}\}/2\Bigr)\nonumber\\
&
&-\sin(2\phi)\Bigl(\cos\theta\{S_{x'},S_{y'}\}+\sin\theta\{S_{y'},S_{z'}\}\Bigr)\Bigr],\nonumber\\
\end{eqnarray}
where $\{A,B\}=AB+BA$ is the anticommutator.

\subsection{Global Hamiltonian
matrix elements}

 The operations of the rotated global anisotropy interactions upon
these states may be written as
\begin{eqnarray}
{\cal
H}_b'|\varphi_s^m\rangle&=&-\frac{J_b}{4}\Bigl[\Bigl(4m^2+2[s(s+1)-3m^2]\sin^2\theta\Bigr)|
\varphi_s^m\rangle\nonumber\\
& &-\sin(2\theta)\sum_{\sigma=\pm1}(2m+\sigma)A_s^{\sigma
m}|\varphi_s^{m+\sigma}\rangle\nonumber\\
& &+\sin^2\theta\sum_{\sigma=\pm1}F_s^{\sigma
m}|\varphi_s^{m+2\sigma}\rangle\Bigr],
\label{Hbprime}\end{eqnarray} and
\begin{eqnarray} {\cal
H}_d'|\varphi_s^m\rangle&=&-\frac{J_d}{4}\Bigl(2\sin^2\theta\cos(2\phi)[3m^2-s(s+1)]|\varphi_s^m\rangle\nonumber\\
& &+2\sin\theta\sum_{\sigma=\pm1}(2m+\sigma)A_s^{\sigma
m}\nonumber\\
& &\times[\cos\theta\cos(2\phi)+i\sigma\sin(2\phi)]|\varphi_s^{m+\sigma}\rangle\nonumber\\
& &+\sum_{\sigma=\pm1}F_s^{\sigma
m}\bigl[(1+\cos^2\theta)\cos(2\phi)\nonumber\\
&
&+2i\sigma\cos\theta\sin(2\phi)\bigr]|\varphi_s^{m+2\sigma}\rangle\Bigr),\label{Hdprime}
\end{eqnarray}
where $F_s^x$ is defined by  Eq. (\ref{Fsx}).  We note that in
this representation, both ${\cal H}_b$ and ${\cal H}_d$ preserve
the global spin quantum number $s$, but allow $\Delta m=\pm1,\pm2$
transitions. As a check on the algebra, we verified that for
$s_1=1/2$, Eq. (\ref{halfeigen}) with $J_a=J_c=0$ is obtained from
Eqs. (\ref{Hbprime}) and (\ref{Hdprime}).

\subsection{Local Hamiltonian}

With regard to the local spin anisotropy terms in the rotated
coordinate system, we write
\begin{eqnarray}
{\cal H}_a'&=&-J_a\Bigl({\cal O}_{1}\cos^2\theta+{\cal
O}_{2}\sin^2\theta-\frac{\sin(2\theta)}{2}{\cal O}_{3}\Bigr)
\end{eqnarray}
and \begin{eqnarray} {\cal
H}_e'&=&-J_e\Bigl[\cos(2\phi)\Bigl({\cal O}_1\sin^2\theta+{\cal
O}_2\cos^2\theta\nonumber\\
& &\qquad+\frac{1}{2}\sin(2\theta){\cal O}_3-{\cal O}_4\Bigr)\nonumber\\
& &-\sin(2\phi)\Bigl({\cal O}_5\cos\theta+ {\cal
O}_6\sin\theta\Bigr)\Bigr],
\end{eqnarray}
where
\begin{eqnarray}
{\cal O}_{1}&=&\sum_{i=1}^2S_{iz'}^2,\label{O1}\\
{\cal O}_{2}&=&\sum_{i=1}^2S_{ix'}^2,\label{O2}\\
{\cal
O}_{3}&=&\sum_{i=1}^2(S_{iz'}S_{ix'}+S_{ix'}S_{iz'}),\label{O3}\\
{\cal O}_4&=&\sum_{i=1}^2S_{iy'}^2,\\
{\cal O}_5&=&\sum_{i=1}^2(S_{ix'}S_{iy'}+S_{iy'}S_{ix'}),
\end{eqnarray}
and\begin{eqnarray}
 {\cal
O}_6&=&\sum_{i=1}^2(S_{iy'}S_{iz'}+S_{iz'}S_{iy'}).
\end{eqnarray}

\subsection{Local Hamiltonian matrix element components}

 The operations of these interactions are
given by
\begin{eqnarray}
{\cal
O}_{1}|\varphi_s^m\rangle&=&\frac{1}{2}\Bigl(G_{s,s_1}^m|\varphi_s^m\rangle\nonumber\\
&
&+\sum_{\sigma'=\pm1}H_{s,s_1}^{m,\sigma'}|\varphi_{s+2\sigma'}^m\rangle\Bigr),\\
{\cal
O}_{2}|\varphi_s^m\rangle&=&\frac{1}{8}\Bigl(M_{s,s_1}^m|\varphi_s^m\rangle-\sum_{\sigma'=\pm1}N_{s,s_1}^{m,\sigma'}|\varphi_{s+2\sigma'}^m\rangle\nonumber\\
& &+\sum_{\sigma=\pm1}L_{s,s_1}^{\sigma m}|\varphi_s^{m+2\sigma}\rangle\nonumber\\
& &+\sum_{\sigma,\sigma'=\pm1}K_{s,s_1}^{\sigma
m,\sigma'}|\varphi_{s+2\sigma'}^{m+2\sigma}\rangle\Bigr),\\
{\cal
O}_{3}|\varphi_s^m\rangle&=&\frac{1}{4}\sum_{\sigma=\pm1}\Bigl(P_{s,s_1}^{m,\sigma}|\varphi_s^{m+\sigma}\rangle\nonumber\\
& & -\sum_{\sigma'=\pm1}\sigma\sigma' R_{s,s_1}^{\sigma
m,\sigma'}|\varphi_{s+2\sigma'}^{m+\sigma}\rangle\Bigr),\\
{\cal
O}_4|\varphi_s^m\rangle&=&\frac{1}{8}\Bigl(M_{s,s_1}^m|\varphi_s^m\rangle-\sum_{\sigma'=\pm1}N_{s,s_1}^{m,\sigma'}|\varphi_{s+2\sigma'}^m\rangle\nonumber\\
& &-\sum_{\sigma=\pm1}L_{s,s_1}^{\sigma m}|\varphi_s^{m+2\sigma}\rangle\nonumber\\
& &-\sum_{\sigma,\sigma'=\pm1}K_{s,s_1}^{\sigma
m,\sigma'}|\varphi_{s+2\sigma'}^{m+2\sigma}\rangle\Bigr),\\
{\cal
O}_5|\varphi_s^m\rangle&=&\frac{1}{4i}\sum_{\sigma=\pm1}\sigma\Bigl(L_{s,s_1}^{\sigma m}|\varphi_s^{m+2\sigma}\rangle\nonumber\\
& &+\sum_{\sigma'=\pm1}K_{s,s_1}^{\sigma
m,\sigma'}|\varphi_{s+2\sigma'}^{m+2\sigma}\rangle\Bigr),\\
{\cal
O}_6|\varphi_s^m\rangle&=&\frac{1}{4i}\sum_{\sigma=\pm1}\Bigl(\sigma P_{s,s_1}^{m,\sigma}|\varphi_s^{m+\sigma}\rangle\nonumber\\
& &-\sum_{\sigma'=\pm1}\sigma' R_{s,s_1}^{\sigma
m,\sigma'}|\varphi_{s+2\sigma'}^{m+\sigma}\rangle\Bigr),
\end{eqnarray}
where \begin{eqnarray} M^m_{s,s_1}&=&-4m^2\alpha_{s,s_1}+4[s(s+1)-1](1-\alpha_{s,s_1}),\nonumber\\
& &\\
 N_{s,s_1}^{m,\sigma'}&=&\sum_{\sigma=\pm1}C_{s+(\sigma'+1)/2,s_1}^{-\sigma\sigma'm-(\sigma'+1)/2}
C_{s+(3\sigma'+1)/2,s_1}^{\sigma\sigma'm+(\sigma'-1)/2},\label{Nss1msigma}\\
P_{s,s_1}^{m,\sigma}&=&2A_s^{\sigma m}(2m+\sigma)\alpha_{s,s_1},\\
\noalign{\rm and}
 R_{s,s_1}^{x,\sigma'}&=&C_{s+(\sigma'+1)/2,s_1}^{-x\sigma'-(\sigma'+1)/2}D_{s+(3\sigma'+1)/2,s_1}^{m+\sigma}\nonumber\\
&
&+C_{s+(3\sigma'+1)/2,s_1}^{-x\sigma'-(\sigma'+1)/2}D_{s+(\sigma'+1)/2,s_1}^m,\label{Rss1xsigma}
\end{eqnarray}  where $\eta_{s,s_1}$, $G_{s,s_1}^m$, $H_{s,s_1}^{m,\sigma'}$,
$K_{s,s_1}^{x,\sigma'}$, and $L_{s,s_1}^x$ are given by Eqs.
(\ref{eta}) and (\ref{Gss1m})-(\ref{Lss1}), respectively.  We note
that for $\sigma'=\pm1$,
$N_{s,s_1}^{m,\sigma'}=2H_{s,s_1}^{m,\sigma'}$.

\section{Appendix C}
\subsection{Second order induction representation Hamiltonian}

In this appendix, we evaluate the corrections to the eigenstate
energies second order in the four anisotropy interaction energies
$J_j$ for $j=a,b,d,e$. The operations of the rotated Hamilatonian
${\cal H}'$ upon the eigenstates $|\varphi_s^m\rangle$ may be
written as
\begin{eqnarray}
{\cal
H}'|\varphi_s^m\rangle&=&(E_{s}^{m,(0)}+E_{s,s_1}^{m,(1)})|\varphi_s^m\rangle
+\sum_{\sigma'=\pm1}{\cal W}_{s,s_1}^{m,\sigma'}|\varphi_{s+2\sigma'}^m\rangle\nonumber\\
& &+\sum_{\sigma=\pm1}\Bigl({\cal
U}_{s,s_1}^{m,\sigma}|\varphi_s^{m+\sigma}\rangle+{\cal
V}_{s,s_1}^{m,\sigma}|\varphi_s^{m+2\sigma}\rangle\Bigr)\nonumber\\
& &+\sum_{\sigma,\sigma'=\pm1}\Bigl({\cal
X}_{s,s_1}^{m,\sigma,\sigma'}|\varphi_{s+2\sigma'}^{m+\sigma}\rangle\nonumber\\
& &\qquad+{\cal
Y}_{s,s_1}^{m,\sigma,\sigma'}|\varphi_{s+2\sigma'}^{m+2\sigma}\rangle\Bigr),\label{Hprime}
\end{eqnarray}
where
\begin{eqnarray}
{\cal
U}_{s,s_1}^{m,\sigma}(\theta,\phi)&=&\frac{1}{4}(2m+\sigma)A_s^{\sigma
m}\Bigl[
\sin(2\theta)\Bigl(\tilde{J}_{b,a}^{s,s_1}\nonumber\\
& &\qquad-\tilde{J}_{d,e}^{s,s_1}\cos(2\phi)\Bigr)\nonumber\\
& &-2i\sigma\tilde{J}_{d,e}^{s,s_1}\sin\theta\sin(2\phi)\Bigr],\\
{\cal V}_{s,s_1}^{m,\sigma}(\theta,\phi)&=&-\frac{1}{4}F_s^{\sigma
m}\Bigl[\tilde{J}_{b,s}^{s,s_1}\sin^2\theta\nonumber\\
& &+\tilde{J}_{d,e}^{s,s_1}(1+\cos^2\theta)\cos(2\phi)\nonumber\\
& &+2i\sigma\tilde{J}_{d,e}^{s,s_1}\cos\theta\sin(2\phi)\Bigr],\\
{\cal
W}_{s,s_1}^{m,\sigma'}(\theta,\phi)&=&-\frac{1}{2}H_{s,s_1}^{m,\sigma'}[J_a\cos^2\theta+J_e\sin^2\theta\cos(2\phi)]\nonumber\\
& &+\frac{1}{8}N_{s,s_1}^{m,\sigma'}\sin^2\theta[J_a-J_e\cos(2\phi)],\\
{\cal
X}_{s,s_1}^{m,\sigma,\sigma'}(\theta,\phi)&=&-\frac{\sigma\sigma'}{8}R_{s,s_1}^{\sigma
m,\sigma'}\Bigl(\sin(2\theta)[J_a-J_e\cos(2\phi)]\nonumber\\
& &\qquad-2i\sigma J_e\sin\theta\sin(2\phi)\Bigr),\\
\noalign{\rm and} & &\nonumber\\ {\cal Y}_{s,s_1}^{\sigma
m,\sigma'}(\theta,\phi)&=&-\frac{1}{8}K_{s,s_1}^{\sigma
m,\sigma'}\Bigl[J_a\sin^2\theta\nonumber\\
& &
\qquad+J_e\Bigl((1+\cos^2\theta)\cos(2\phi)\nonumber\\
& &\qquad+2i\sigma\cos\theta\sin(2\phi)\Bigr)\Bigr],
\end{eqnarray}
where $E_{s,s_1}^{m,(0)}$ and $E_{s,s_1}^{m,(1)}$ are given by
Eqs. (\ref{E0}) and (\ref{Esm1}), respectively,
$H_{s,s_1}^{m,\sigma'}$ and $K_{s,s_1}^{x,\sigma'}$ are given by
Eqs. (\ref{Hss1msigma}) and (\ref{Kss1xsigma}), respectively,
$N_{s,s_1}^{m,\sigma'}$ and $R_{s,s_1}^{x,\sigma'}$ are given by
Eqs. (\ref{Nss1msigma}) and (\ref{Rss1xsigma}), respectively, and
$\tilde{J}_{b,a}^{s,s_1}$ and $\tilde{J}_{d,e}^{s,s_1}$ are given
by Eqs. (\ref{tildeJba}) and (\ref{tildeJde}), respectively.

\subsection{Second order eigenstate energies}

 From Eq. (\ref{Hprime}), the second order
eigenstate energies may be written as
\begin{eqnarray}
E_{s,s_1}^{m,(2)}&=&\frac{1}{\gamma
B}\sum_{\sigma=\pm1}\sigma\Bigl(\Bigl|{\cal
U}_{s,s_1}^{m,\sigma}\Bigr|^2+\frac{1}{2}\Bigl|{\cal
V}_{s,s_1}^{m,\sigma}\Bigr|^2\Bigr)\nonumber\\
& &+\sum_{\sigma'=\pm1}\frac{|{\cal
W}_{s,s_1}^{m,\sigma'}|^2}{J[2+(2s+1)\sigma']}\nonumber\\
& &+\sum_{\sigma,\sigma'\pm1}\Bigl(\frac{|{\cal
X}_{s,s_1}^{m,\sigma,\sigma'}|^2}{J[2+(2s+1)\sigma']+\sigma\gamma
B}\nonumber\\
& &+\frac{|{\cal
Y}_{s,s_1}^{m,\sigma,\sigma'}|^2}{J[2+(2s+1)\sigma']+2\sigma\gamma
B}\Bigr).
\end{eqnarray}

For simplicity, we rewrite this as

\begin{eqnarray}
E_{s,s_1}^{m,(2)}&=&E_{s,s_1}^{m,(2){\cal
U}}+E_{s,s_1}^{m,(2){\cal
V}}+E_{s,s_1}^{m,(2){\cal W}}\nonumber\\
& &+E_{s,s_1}^{m,(2){\cal X}}+E_{s,s_1}^{m,(2){\cal Y}},\\
E_{s,s_1}^{m,(2){\cal U}}&=&\frac{m\sin^2\theta}{2\gamma
B}[4s(s+1)-8m^2-1]\nonumber\\
&
&\times\Bigl(\cos^2\theta[\tilde{J}_{b,a}^{s,s_1}-\cos(2\phi)\tilde{J}_{d,e}^{s,s_1}]^2\nonumber\\
& &\qquad+\sin^2(2\phi)(\tilde{J}_{d,e}^{s,s_1})^2\Bigr),\label{Esm2}\\
E_{s,s_1}^{m,(2){\cal V}}&=&-\frac{m}{8\gamma B}[2s(s+1)-2m^2-1]\nonumber\\
&
&\times\Bigl([\sin^2\theta\tilde{J}_{b,a}^{s,s_1}+(1+\cos^2\theta)\cos(2\phi)\tilde{J}_{d,e}^{s,s_1}]^2\nonumber\\
&
&\qquad+4\cos^2\theta\sin^2(2\phi)(\tilde{J}_{d,e}^{s,s_1})^2\Bigr),\\
E_{s,s_1}^{m,(2){\cal
W}}&=&\frac{d_{s,s_1}^m}{16J}\Bigl(J_a-3[J_a\cos^2\theta+J_e\sin^2\theta\cos(2\phi)]\Bigr)^2,\nonumber\\
& &\label{Esm2W}\\
E_{s,s_1}^{m,(2){\cal
X}}&=&\frac{f_{s,s_1}^m(\gamma B/J)\sin^2\theta}{2J}\nonumber\\
& &\times\Bigl([J_a-J_e\cos(2\phi)]^2\cos^2\theta\nonumber\\
& &\qquad+\sin^2(2\phi)J_e^2\Bigr),\label{Esm2X}\\
E_{s,s_1}^{m,(2){\cal
Y}}&=&\frac{g_{s,s_1}^m(\gamma B/J)}{64J}\Bigl[\Bigl(J_a\sin^2\theta\nonumber\\
& &\qquad+J_e(1+\cos^2\theta)\cos(2\phi)\Bigr)^2\nonumber\\
& &\qquad+4\cos^2\theta\sin^2(2\phi)J_e^2\Bigr],\label{Esm2Y}
\end{eqnarray}
where
\begin{eqnarray}
d_{s,s_1}^m&=&-\frac{(s^2-m^2)[(s-1)^2-m^2]\eta_{s,s_1}^2\eta_{s-1,s_1}^2}{(2s-1)}\nonumber\\
&&+\eta_{s+2,s_1}^2\eta_{s+1,s_1}^2\nonumber\\
& &\times
\frac{[(s+1)^2-m^2][(s+2)^2-m^2]}{(2s+3)},\nonumber\\
f_{s,s_1}^m(x)&=&-\frac{\eta_{s,s_1}^2\eta_{s-1,s_1}^2(s^2-m^2)}{(2s-1)^2-x^2}\nonumber\\
& &\times\Bigl((2s-1)[(s-1)(s-2)+m^2]\nonumber\\
& &\qquad-m(2s-3)x\Bigr)\nonumber\\
&
&+\frac{\eta_{s+2,s_1}^2\eta_{s+1,s_1}^2[(s+1)^2-m^2]}{(2s+3)^2-x^2}\nonumber\\
& &\times\Bigl((2s+3)[(s+2)(s+3)+m^2]\nonumber\\
& &\qquad-m(2s+5)x\Bigr),\\
g_{s,s_1}^m(x)&=&\frac{2\eta_{s,s_1}^2\eta_{s-1,s_1}^2}{(2s-1)^2-4x^2}\Bigl((2s-1)[m^4\nonumber\\
& &+m^2(6s^2-18s+11)\nonumber\\& &+s(s-1)(s-2)(s-3)]\nonumber\\
& &-4mx(2s-3)(s^2-3s+1+m^2)\Bigr)\\
& &
-\frac{2\eta_{s+2,s_1}^2\eta_{s+1,s_1}^2}{(2s+3)^2-4x^2}\Bigl((2s+3)[m^4\nonumber\\
& &+m^2(6s^2+29s+34)\nonumber\\
& &+(s+1)(s+2)(s+3)(s+4)]\nonumber\\
& &-4mx(2s+5)(s^2+5s+5+m^2)\Bigr).
\end{eqnarray}

There is a remarkable amount of symmetry in the angular dependence
of the eigenstate energies.  We note that $E_{s,s_1}^{m,(2){\cal
X}}(\theta,\phi)$ and $E_{s,s_1}^{m,(2){\cal U}}(\theta,\phi)$
have the same forms, differing in the replacements of the
interactions $\tilde{J}_{b,a}^{s,s_1}$ and
$\tilde{J}_{d,e}^{s,s_1}$ with $J_a$ and $J_e$, respectively, and
with  different overall constant functions. The same comparison
can also be made with $E_{s,s_1}^{m,(2){\cal Y}}(\theta,\phi)$ and
$E_{s,s_1}^{m,(2){\cal V}}(\theta,\phi)$.  In addition, we note
that there is a remarkable similarity in the $\theta,\phi$
dependence of $E_{s,s_1}^{m,(2){\cal W}}$ with that of the local
spin anisotropy part of $B_{s,s_1}^{{\rm lc}(1)}(\theta,\phi)$ in
Eq. (\ref{Bstep}).

\section{Appendix D}

The contributions to the $s$th level crossing second order in the
anisotropy interactions are calculated as indicated in Eq.
(\ref{Bstep2}), and are found to be
\begin{eqnarray}\Bigl(E_{s,s_1}^{s,(2)}-E_{s-1,s_1}^{s-1,(2)}\Bigr)\Bigr|_{B=-Js/\gamma}&=&\sum_{n=1}^7a_n(s,s_1)f_n(\theta,\phi).\label{level2}\nonumber\\
\end{eqnarray}

\subsection{Second order level crossing angular functions}

\begin{eqnarray}
f_1(\theta,\phi)&=&\frac{\sin^2\theta}{J}\Bigl([J_b-\cos(2\phi)J_d]^2\cos^2\theta\nonumber\\
& &\qquad\qquad+J_d^2\sin^2(2\phi)\Bigr),\\
f_2(\theta,\phi)&=&\frac{\sin^2\theta}{J}\Bigl([J_b-J_d\cos(2\phi)][J_a-J_e\cos(2\phi)]\nonumber\\
& &\qquad\times\cos^2\theta  +J_dJ_e\sin^2(2\phi)]\Bigr),\\
f_3(\theta,\phi)&=&\frac{\sin^2\theta}{J}\Bigl([J_a-\cos(2\phi)J_e]^2\cos^2\theta\nonumber\\
& &\qquad\qquad+J_e^2\sin^2(2\phi)\Bigr),\\
f_4(\theta,\phi)&=&\frac{1}{J}\Bigl(J_b\sin^2\theta+J_d(1+\cos^2\theta)\cos(2\phi)\Bigr)^2\nonumber\\
& &+4\frac{J_d^2}{J}\cos^2\theta\sin^2(2\phi),\\
f_5(\theta,\phi)&=&\frac{1}{J}[J_b\sin^2\theta+J_d(1+\cos^2\theta)\cos(2\phi)]\nonumber\\
&
&\times[J_a\sin^2\theta+J_e(1+\cos^2\theta)\cos(2\phi)]\nonumber\\
& &+4\frac{J_dJ_e}{J}\cos^2\theta\sin^2(2\phi),\\
f_6(\theta,\phi)&=&\frac{1}{J}\Bigl(J_a\sin^2\theta+J_e(1+\cos^2\theta)\cos(2\phi)\Bigr)^2\nonumber\\
& &+4\frac{J_e^2}{J}\cos^2\theta\sin^2(2\phi),\\
f_7(\theta,\phi)&=&\frac{1}{J}\Bigl(J_a-3[J_a\cos^2\theta+J_e\sin^2\theta\cos(2\phi)]\Bigr)^2.\nonumber\\
\end{eqnarray}

\subsection{Second order level crossing coefficients}

The coefficients are given by
\begin{eqnarray}
a_1(s,s_1)&=&-\frac{(8s-9)}{2s},\\
a_2(s,s_1)&=&-\frac{[a_{2,0}(s)+4s_1(s_1+1)a_{2,1}(s)]}{s(2s+1)(2s+3)},\\
a_{2,0}(s)&=&-3(9+s-23s^2+4s^3+12s^4),\nonumber\\
a_{2,1}(s)&=&-9+8s+4s^2,\\
a_3(s,s_1)&=&a_3^{\cal U}(s,s_1)+a_3^{\cal X}(s,s_1),\\
a_3^{\cal
U}(s,s_1)&=&\frac{[3-3s-3s^2+4s_1(s_1+1)]^2}{2(2s+3)^2}\nonumber\\
& &-\frac{(s-1)[3+3s-3s^2+4s_1(s_1+1)]^2}{2s(2s+1)^2},\nonumber\\
& &\\
 a_3^{\cal
X}(s,s_1)&=&\frac{[s(s+2)-4s_1(s_1+1)]}{2(s+1)(s+3)(2s+1)^2(2s+3)^2}\nonumber\\
& &\times\frac{a_{3,0}^{\cal X}(s)+4s_1(s_1+1)a_{3,1}^{\cal X}(s)}{(2s+5)(3s+1)},\nonumber\\
a_{3,0}^{\cal X}(s)&=&(s+1)(s+3)(51+114s\nonumber\\
& &+209s^2+302s^3+164s^4+24s^5),\\
a_{3,1}^{\cal X}(s)&=&129+318s+395s^2\nonumber\\
& &+442s^3+300s^4+72s^5,\nonumber\\
a_{4}(s,s_1)&=&\frac{(4s-3)}{8s},\\
a_5(s,s_1)&=&-\frac{3[a_{5,0}(s)+4s_1(s_1+1)]}{4s(2s+1)(2s+3)},\\
a_{5,0}(s)&=&3+3s-5s^2-4s^3,\\
a_6(s,s_1)&=&a_6^{\cal V}(s,s_1)+a_6^{\cal Y}(s,s_1),\\
 a_6^{\cal
 V}(s,s_1)&=&\frac{[3-3s-3s^2+4s_1(s_1+1)]^2}{8(2s-1)(2s+3)^2}\nonumber\\
 &
 &-\frac{(s-1)[3+3s-3s^2+4s_1(s_1+1)]^2}{8s(2s-3)(2s+1)^2},\nonumber\\
 & &\\
 a_6^{\cal
Y}(s,s_1)&=&\frac{-1}{96(2s-3)(2s-1)(2s+1)^2(2s+3)^2}\nonumber\\
&
&\times\frac{\sum_{n=0}^2a_{6,n}^{\cal Y}(s)[4s_1(s_1+1)]^n}{(2s+5)(4s+1)(4s+3)},\nonumber\\
a_{6,0}^{\cal Y}(s)&=&s\Bigl(-5184+20898s+123003s^2 \nonumber\\
& &+81669s^3-304665s^4-521611 s^5\nonumber\\
& &-168482s^6+163132s^7+124888s^8\nonumber\\
& &+12008s^9+256s^{10}+128s^{11}\Bigr),\\
a_{6,1}^{\cal Y}(s)&=&5832+6939s+855s^2+17694s^3\nonumber\\
& &+22242s^4-1548s^5-3464s^6\nonumber\\
& &+3728s^7+1920s^8-512s^9-256s^{10},\\
a_{6,2}^{\cal Y}(s)&=&4536+5787s-3111s^2+4782s^3\nonumber\\
& &+16972s^4+5144s^5-864s^6\nonumber\\
& &+256s^7+128s^8,\\
a_7(s,s_1)&=&\frac{[s(s+2)-4s_1(s_1+1)]}{4(2s+1)^3(2s+3)^3(2s+5)}\nonumber\\
&
&\times\Bigl(4s_1(s_1+1)(-1+38s+24s^2+60s^3)\nonumber\\
& &+(s+1)(3+67s+94s^2+44s^3+8s^4)\Bigr).\nonumber\\
\end{eqnarray}

By expanding the solutions in the crystal representation to second
order in the $J_j$, we have explicitly checked these expressions
for $s_1=1/2, s=1$, and for $s_1=1$, $s=1,2$.  We note that for
$s_1=1/2$, only $a_1$ and $a_4$ are non-vanishing.


\begin{thebibliography}{99}
\bibitem{background}
R. Sessoli, D. Gatteschi, A. Caneschi, and M. A. Novak, Nature
(London) {\bf 365}, 141 (1993).

\bibitem{sarachik}
J. R. Friedman, M. P. Sarachik, J. Tejada, and R. Ziolo, Phys.
Rev. Lett. {\bf 76}, 3830 (1996).

\bibitem{loss}
M. N. Leuenberger and D. Loss, Nature (London) {\bf 410}, 789
(2001).

\bibitem{Fe8}
W. Wernsdorfer, T. Ohm, C. Sangregorio, R. Sessoli, D. Mailly, and
C. Paulsen, Phys. Rev. Lett. {\bf 82}, 3903 (1999).

\bibitem{WS}
W. Wernsdorfer and R. Sessoli, Science {\bf 284}, 133 (1999).

\bibitem{Fe8spin9}
D. Zipse, J. M. North, N. S. Dalal, S. Hill, and R. S. Edwards,
Phys. Rev. B {\bf 68}, 184408 (2003).

\bibitem{Fering1}
M. Affronte, A. Cornia, A. Lascialfari, F. Borsa, D. Gatteschi, J.
Hinderer, M. Horvati{\'c}, A. G. M. Jansen, and M.-H. Julien,
Phys. Rev. Lett. {\bf 88}, 167201 (2002).

\bibitem{Fe6ring}
O. Waldmann, J. Sch{\"u}lein, R. Koch, P. M{\"u}ller, I. Bernt, R.
W. Saalfrank, H. P. Andres, H. U. G{\"u}del, and P. Allenspach,
Inorg. Chem. {\bf 38},  5879 (1999).

\bibitem{Fe8ring}
O. Waldmann, R. Koch, S. Schromm, J. Sch{\"u}lein, P. M{\"u}ller,
I. Bernt, R. W. Saalfrank, F. Hempel, and E. Balthes, Inorg. Chem.
{\bf 40},  2986 (2001).

\bibitem{Fering2}
H. Nakano and S. Miyashita, J. Phys. Chem. Solids {\bf 63}, 1519
(2002).

\bibitem{V2neutron}
D. A. Tennant, S. E. Nagler, A. W. Garrett, T. Barnes, and C. C.
Torardi, Phys. Rev. Lett. {\bf 78}, 4998 (1997).

\bibitem{Gudel}
 H. U. G{\"u}del, Neutron News {\bf 7}, 24 (1996)

\bibitem{V2P2O9}
A. W. Garrett, S. E. Nagler, D. A. Tennant, B. C. Sales, and T.
Barnes, Phys. Rev. Lett. {\bf 79}, 745 (1997).

\bibitem{ek}
D. V. Efremov and R. A. Klemm, Phys. Rev. B {\bf 66}, 174427
(2002).

\bibitem{Fe2}
F. Le Gall, F. Fabrizi de Biani, A. Caneschi, P. Cinelli, A.
Cornia, A. C. Fabretti, and D. Gatteschi, Inorg. Chim. Acta {\bf
262}, 123 (1997).

\bibitem{Fe2mag}
Y. Shapira, M. T. Liu, S. Foner, R. J. Howard, and W. H.
Armstrong, Phys. Rev. B {\bf 63}, 094422 (2001).

\bibitem{Fe2Cl}
Y. Shapira, M. T. Liu, S. Foner, C. E. Dub{\'e}, and P. J.
Bonitatebus, Jr., Phys. Rev. B {\bf 59}, 1046 (1999).


\bibitem{taft}
K. L. Taft, C. D. Delfs, G. C. Papaefthymiou, S. Foner, D.
Gatteschi, and S. J. Lippard, J. Am. Chem. Soc. {\bf 116}, 823
(1994).

\bibitem{Fe2Cl3}
J. D. Walker and R. Poli, Inorg. Chem. {\bf 29}, 756 (1990).

\bibitem{Fe2Clnew}
J. A. Bertrand, J. L. Breece, and P. G. Eller, Inorg. Chem. {\bf
13}, 125 (1974).


\bibitem{Mn4dimer}
R. Tiron, W. Wernsdorfer, D. Foquet-Albiol, N. Aliaga-Alcalde, and
G. Christou, Phys. Rev. Lett. {\bf 91}, 227203 (2003).


\bibitem{Mn4dimerDalal}
J. M. North, N. S. Dalal, D. Foquet-Albiol, A. Vinslava, and G.
Christou, Phys. Rev. B {\bf 69}, 174419 (2004).


\bibitem{WaldmannNi}
O. Waldmann, J. Hassmann, P. M{\"u}ller, D. Volkmer, U. S.
Schubert, and J.-M. Lehn, Phys. Rev. B {\bf 58}, 3277 (1998).

\bibitem{Almenar}
J. J. Borr{\'a}s-Almenar, J. M. Clemente-Juan, E. coronado, and B.
S. Tsukerblat, Inorg. Chem. {\bf 38},  6081 (1999).


\bibitem{klemm}
R. A. Klemm and J. R. Clem, Phys. Rev. B {\bf 21}, 1868 (1980); R.
A. Klemm, SIAM J. Appl. Math. {\bf 55}, 986 (1995).

\bibitem{Goldstein}
H. Goldstein, {\it Classical Mechanics}, (Addison-Wesley, Reading,
MA 1965), p. 109.


\end{thebibliography}
\end{document}